\documentclass[apj]{emulateapj}

\def\gtorder{\mathrel{\raise.3ex\hbox{$>$}\mkern-14mu
             \lower0.6ex\hbox{$\sim$}}}
\def\ltorder{\mathrel{\raise.3ex\hbox{$<$}\mkern-14mu
             \lower0.6ex\hbox{$\sim$}}}

\slugcomment{Draft of \today}

\shorttitle{Radio transients}

\shortauthors{Ofek et al.}

\begin{document}

\title{Long duration radio transients lacking optical counterparts are possibly Galactic Neutron Stars}
\author{E.~O.~Ofek\altaffilmark{1}$^{,}$\altaffilmark{2},
B.~Breslauer\altaffilmark{3}$^{,}$\altaffilmark{4},
A.~Gal-Yam\altaffilmark{5},
D.~Frail\altaffilmark{4},
M.~M.~Kasliwal\altaffilmark{1},
S.~R.~Kulkarni\altaffilmark{1} \&
E.~Waxman\altaffilmark{5}}

\altaffiltext{1}{Division of Physics, Mathematics and Astronomy, California Institute of Technology, Pasadena, CA 91125, USA}
\altaffiltext{2}{Einstein fellow}
\altaffiltext{3}{Department of Physics and Astronomy, Oberlin College, Oberlin, Ohio 44074-1088}
\altaffiltext{4}{National Radio Astronomy Observatory, P.O. Box O, Socorro, NM 87801}
\altaffiltext{5}{Benoziyo Center for Astrophysics, Weizmann Institute of Science, 76100 Rehovot, Israel}

\begin{abstract}


Recently, a new class of radio transients in the 5-GHz band and
with durations of the order of hours
to days, lacking any visible-light counterparts, was detected by Bower
and collaborators.
We present new deep near-Infrared (IR) observations of the field containing
these transients, and find no counterparts down to a limiting magnitude of
$K=20.4$\,mag. We argue that the bright ($>1$\,Jy) radio transients recently
reported by Kida et al. are consistent with being additional examples of the
Bower et al. transients. We refer to these groups of events as
``long-duration radio transients''. The main characteristics of this population
are: time scales longer than 30\,minute but shorter than several days;
very large rate, $\sim10^{3}$\,deg$^{-2}$\,yr$^{-1}$;
progenitors sky surface density of $>60$\,deg$^{-2}$ (at $95\%$ confidence)
at Galactic latitude $\sim40^{\circ}$;
1.4--5\,GHz spectral slopes, $f_{\nu}\propto \nu^{\alpha}$, with $\alpha\gtorder0$;
and most notably the lack of any X-ray, visible-light, near-IR, and radio counterparts
in quiescence. We discuss putative known astrophysical objects that may be
related to these transients and rule out an association with many types of
objects including supernovae, gamma-ray bursts, quasars, pulsars, and
M-dwarf flare stars. Galactic brown-dwarfs or some sort
of exotic explosions in the intergalactic medium remain plausible
(though speculative) options. We argue that an attractive progenitor candidate
for these radio transients is the class of Galactic isolated old neutron stars (NS).
We confront this hypothesis with Monte-Carlo simulations of the space
distribution of old NSs, and
find satisfactory agreement for the large areal density.
Furthermore,
the lack of quiescent counterparts is explained quite naturally.
In this framework we find:
the mean distance to events in the Bower et al. sample is of order kpc;
the typical distance to the Kida et al. transients are constrained to be between 30\,pc
and 900\,pc (at the 95\% confidence level);
these events should repeat with a time scale of order several months;
and sub-mJy level bursts should exhibit Galactic latitude dependence.
We discuss two possible mechanisms giving rise to the observed radio
emission:
incoherent synchrotron emission and coherent emission.
We speculate that if the latter is correct,
the long duration radio transients are sputtering ancient
pulsars or magnetars and will
exhibit pulsed emission.

\end{abstract}


\section{Introduction}
\label{sec:Introduction}

Large field-of-view
radio telescope facilities
such as the Parks multi-beam facility,
the Arecibo multi-beam instrument,
the Allen Telescope Array (DeBoer et al. 2004),
and the Low Frequency Array (Falcke et al. 2007)
have reinvigorated 
the radio-frequency time domain frontier.
First signs of this ``revolution'' are indicated by the
discoveries of new classes of radio transients.
Examples include: several Galactic center sources
(e.g., Hyman  et al. 2005; 2009);
Rotating Radio Anomalous Transients (RRATs; McLaughlin et al. 2006)
which represent a previously unknown class of radio pulsars,
probably twice as abundant as their ``normal'' cousins;
and the powerful ($\sim30$~Jy) radio ``Sparker'',
with a time scale of several milliseconds
(Lorimer et al. 2007; Kulkarni et al. 2009).

Here, we focus on yet another emerging class of
mysterious radio transients.
In a novel approach, Bower et al. (2007)
re-analyzed 944 epochs of Very Large Array\footnote{The Very Large Array is operated
by the National Radio Astronomy Observatory, a facility of the National
Science Foundation operated under cooperative agreement by
Associated Universities, Inc.} (VLA) observations,
taken about once per week for twenty two years,
of a single calibration field.
These authors discovered a total of ten transients,
nine in the 5-GHz band and one in the 8-GHz band.
These transients can be divided into two groups:
``single-epoch'' and ``multi-epoch'' transients.
The eight single-epoch transients, as can be gathered by
their names, were detected in only one epoch.
Given a single epoch detection, one can
only constrain the duration of the transient
by the epochs preceding and succeeding the time at which
the transient was detected (approximately one week).
The lower limit could be as small as the typical integration
time (about 20\,minute).
The two multi-epoch transients were detected after averaging over
two months of data. Thus, the duration of these two events
can be taken to be about two months.

Bower et al. (2007)
split the data for each epoch, consisting of 20 minutes, into
five segments and looked for variability on four-minute time scale.
They did~not find any evidence for variability on
these time scales.
However, the total S/N of
these detections was $\ltorder7$, and therefore
the limits on variability within these 20-minute windows
are appropriately weak.
More importantly, the authors find the circular
polarization is less than $\sim30\%$.

Separately, Kuniyoshi et al. (2006),
Niinuma et al. (2007), and Kida et al. (2008) reported
on a search for radio transients using
an East-West interferometer of the 
Nasu Pulsar Observatory (located in Tochigi Prefecture, Japan)
of Waseda University.
The program consists of daily drift scanning
of the sky towards the local zenith.
These authors reported six bright radio transients
(and several other were mentioned but without details),
with flux density above 1\,Jy
in the 1.4-GHz band.
Five were single epoch transients 
while one was detected on two successive days
with flux densities of 1.7 and 3.2\,Jy, respectively (Niinuma et al. 2007).
In each epoch the transients were detected for about 4\,minutes,
which is the drift scanning time,
and did~not exhibit any significant variation within
the observation.
Unfortunately, these events are not well localized
and have positional uncertainties of the order
of $0.4^{\circ}$ in declination,
and $0.04^{\circ}$ in right ascension.
Kida et al. (2008) stated that the 2-$\sigma$ upper limit
for the rate of these transients is
$0.0049$\,deg$^{-2}$\,yr$^{-1}$.
Arguably, 
these events also have time scales somewhere
in the range of a few minutes to a few days.
Therefore, later we consider the framework in which
both the VLA and the Nasu events have a common origin.

Another relevant survey was conducted by
Levinson et al. (2002) who
compared the
NRAO VLA Sky Survey (NVSS; Condon et al. 1998)
and the ``Faint Images of the Radio Sky at Twenty centimeters''
survey (FIRST; Becker et al. 1995; White et al. 1997).
Both these surveys were undertaken in the 1.4-GHz band
and their 5-$\sigma$ limits
are 3.5 and 1\,mJy, respectively.
%
Levinson et al. (2002) identified nine radio transient candidates
with flux densities greater than 6\,mJy.
Followup observations of these radio transients (Gal-Yam et al. 2006)
showed that seven were spurious and the remaining two were plausible
transients\footnote{The term ``transient'' is used here in the sense that we do~not
detected emission in quiescence.}:
an optically extincted SN in NGC~4216
from which the radio emission lasted for several years; and VLA~J172059.9$+$385229
(discussed in \S\ref{sec:VLA1720}).

Bower et al. (2007) obtained deep visible light images
of their transients.
They found that the multi-epoch event
RT\footnote{Here RT stands for radio transient and the succeeding eight digits are yyyymmdd
where yyyy is the year, mm is the month and dd is the day.}\,19870422
was 1\farcs5 from a $z=0.249$,
$R=20.2$\,mag galaxy.
The peak luminosity of RT\,19870422,
assuming that the transient is related to this galaxy,
is 
consistent (to an order of a magnitude) with
an energetic supernova (SN)
similar to SN\,1998bw (GRB\,980425; Kulkarni et al. 1998)
and SN\,2006aj (Soderberg et al. 2006).
We note that the rate of these events is marginally
consistent with the rate of low-luminosity GRBs
derived by Soderberg et al. (2006).
In contrast, the other multi-epoch transient
RT\,20010331 has no optical counterpart to
a limiting magnitude of $g\approx27.6$\,mag,
$R\approx26.5$\,mag, and $K\approx19.2$\,mag (this paper)
to within $5''$ of the radio source.
The great offset between a putative host galaxy and the radio
transient make this an unusual source.
We note that Cenko et al. (2008) presented
an example of a GRB in a galaxy halo environment.
However, only about $1\%$ of all GRBs occures in
such environments.

The single-epoch radio transient RT\,19840613 falls within
the optical boundary of a $z=0.040$ spiral galaxy, but clearly
lying outside the nucleus of the galaxy.
On the basis of the radio luminosity
(assuming association with the galaxy) and
the nature of the putative host galaxy
this transient is consistent with an origin
similar to that of RT\,19870422 (i.e., a low luminosity GRB).

The remaining seven single-epoch transients
(one detected in 8\,GHz and six at 5\,GHz)
do not have
astrometrically coincident optical,
near-IR, or radio counterparts
and have a point source appearance (see Table~\ref{tab:FastTran}).
This is a major clue in that a large fraction
of gamma ray bursts (GRBs) and most SNe have detectable optical host
galaxies at $R\sim26$\,mag level
(e.g., Ovaldsen et al. 2007).


%
In order to separate the events discussed above from Sparkers
(Lorimer et al. 2007; Kulkarni et al. 2009),
which have very short time scales,
we refer to these events
as ``long-duration radio transients''.

In Table~\ref{tab:FastTran} we summarize
the observational properties of all the long-duration radio
transients.
We define this class as events with no optical identification
and durations between hours to days.
This group include the seven known sources
from Bower et al. (2007)
that are not associated with any optical counterpart
and have time scales shorter than about one week;
RT\,19870422 (see above);
and the six bright transients reported by Kida et al. (2008).

The structure of this paper is as follows.
In \S\ref{sec:VLA1720} we re-examine the case of VLA\,J172059.9$+$385229
(Levinson et al. 2002)
and show it is a spurious event.
In \S\ref{sec:Obs} we present new near-IR observations
of the Bower et al. field.
In \S\ref{sec:Prop} we review the 
basic properties (areal density, annual rate) of the 
long-duration radio transients.
Next, in \S\ref{sec:NotProgenitor} we use the observational clues
to refute several plausible explanations
regarding the nature of the long-duration radio transients.
%
In \S\ref{sec:IONS} we argue
that the most attractive explanation
is that these radio
transients are associated with Galactic isolated old Neutron Stars (NS).
%
Finally, we discuss and summarize the results in \S\ref{sec:Disc}.

%
\begin{deluxetable*}{llllllllllllll}
\tablecolumns{14}
\tablewidth{0pt}
\tablecaption{List of candidate long-duration radio transients}
\tablehead{
\multicolumn{9}{c}{Transient} &
\multicolumn{4}{c}{Lim. mag.} &
\colhead{} \\
\colhead{Epoch} &
\colhead{R.A.} &
\colhead{Dec.} &
\colhead{Band} &
\colhead{$S$} &
\colhead{$\delta{t}$\tablenotemark{a}} &
\colhead{$S_{next}$\tablenotemark{b}} &
\colhead{$S_{deep}$\tablenotemark{c}} &
\colhead{$S/S_{deep}$} &
\colhead{X\tablenotemark{d}} &
\colhead{$g$} &
\colhead{$R$} &
\colhead{$K$} &
\colhead{Ref.} \\
\colhead{}        &
\colhead{J2000}        &
\colhead{J2000}        &
\colhead{GHz}     &
\colhead{$\mu$Jy} &
\colhead{day}     &
\colhead{$\mu$Jy} &
\colhead{$\mu$Jy} &
\colhead{}     &
\colhead{cts}  &
\colhead{mag}  &
\colhead{mag}  &
\colhead{mag}  &
\colhead{} 
}
\startdata
1984 05 02 & 15 02 24.61& $+$78 16 10.1 & 5.0 & $448\pm74$   & 7  & $-10\pm68$   & $<8$  & 56  & 0.08 & 27.6 & 26.5 & 20.4 & 1\\
1986 01 15 & 15 02 26.40& $+$78 17 32.4 & 5.0 & $370\pm67$   & 7  & $199\pm121$  & $<8$  & 46  & 0.07 & 27.6 & 26.5 & 20.4 & 1\\
1986 01 22 & 15 00 50.15& $+$78 15 39.4 & 5.0 & $1586\pm248$ & 7  & $-59\pm164$  & $<15$ & 106 & 0.08 & 27.6 & 26.5 & 20.2 & 1\\
1992 08 26 & 15 02 59.89& $+$78 16 10.8 & 5.0 & $642\pm101$  & 56 & $37\pm83$    & $<9$  & 71  & 0.08 & 27.6 & 26.5 & 19.6 & 1\\
1997 05 28\tablenotemark{e} & 15 00 23.55& $+$78 13 01.4 & 5.0 & $1731\pm232$ & 7  & $90\pm206$   & $<36$ & 48  & 0.07 & 27.6 & 26.5 & 20.0 & 1\\
1999 05 04 & 14 59 46.42& $+$78 20 29.0 & 5.0 & $7042\pm963$ & 21 & $-313\pm1020$& $<117$& 60  & 0.05 & 27.6 & 26.5 & 19.2 & 1\\
\hline
1997 02 05 & 15 01 29.35& $+$78 19 49.2 & 8.4 & $2234\pm288$ & 5  & $857\pm323$  & $<646$& 3.5 & 0.08 & 27.6 & 26.5 & 20.4 & 1\\
\hline
2005 01 10 & 04 45 17   & $+$41 30      & 1.4 & $1.8\times10^{6}$ & 1 & $\ltorder3\times10^{5}$ & $\ltorder9\times10^{3}$ & 200 & 0.02 & & & & 2 \\
2005 03 27 & 06 45 15   & $+$32 00      & 1.4 & $1.2\times10^{6}$ & 1 & $\ltorder3\times10^{5}$ & $\ltorder6\times10^{3}$ & 200 & 0.02 & & & & 2 \\
2005 03 04 & 10 39 43   & $+$32 00      & 1.4 & $1.7\times10^{6}$ & 1 & $\ltorder3\times10^{5}$ & $\ltorder7\times10^{3}$ & 240 & 0.04 & & & & 2 \\
2005 01 02 & 10 43 06   & $+$41 00      & 1.4 & $1.7\times10^{6}$ & 1 & $\ltorder3\times10^{5}$ & $\ltorder1\times10^{3}$ & 1700& 0.02 & & & & 2 \\
2005 02 13\tablenotemark{f} & 14 43 22   & $+$34 39      & 1.4 & $3.2\times10^{6}$ & 1 & $\ltorder3\times10^{5}$ & $\ltorder2\times10^{3}$ & 1600 & 0.02 & & & & 2 \\
2004 03 20 & 17 37 17   & $+$38 08      & 1.4 & $1.0\times10^{6}$ & 1 & $\ltorder3\times10^{5}$ & $\ltorder6\times10^{3}$ & 170 & 0.02 & & & & 2 \\
\hline
\hline
2001 10 31\tablenotemark{g} & 15 03 46.18& $+$78 15 41.7 & 4.0 & $697\pm94$   & 59 & $85\pm85$    & $<37$ & 19  & 0.08 & 27.6 & 26.5 & 19.2 & 1
\enddata
\tablenotetext{a}{Time to next observation.}
\tablenotetext{b}{Flux limit in the next observation.}
\tablenotetext{c}{
Specific flux limit on radio emission at quiescence.
For the Bower et al. transients this limit is obtained from the non detection in the combined image of the Bower et al. field. For the Kida et al. (2008) transients we list the flux of the brightest FIRST or NVSS radio source (or detection limit if no source) in the transient positional error region.}
\tablenotetext{d}{ROSAT 3-$\sigma$ upper limit (in count\,s$^{-1}$) in the 0.12-2.48\,keV band. For the Kida et al. (2008) transients we list the count rate of the brightest ROSAT source (or detection limit if no source) in the transient positional error region. We obtained these limits by calculating
the 3-$\sigma$ noise level due to the background in the
ROSAT X-ray images at the location
of each transient.}
\tablenotetext{e}{A galaxy with $R=19.6$\,mag and $z=0.245$, $5''$ away, probably due to chance coincidence.}
\tablenotetext{f}{Detected on two epochs, separated by one day, with fluxes of 1.7\,Jy and 3.2\,Jy on the first and second epochs, respectively.}
\tablenotetext{g}{RT\,20011031 had a time scale of two~months and is listed here for completeness.}
\tablecomments{A list of candidate long-duration radio transients and their properties.
Transients detected in different frequencies or instruments are separated
by horizontal lines.
The first (second) block lists the six (one) single-epoch transients
detected by Bower et al. (2007) at 5\,GHz (8\,GHz) with no optical counterpart.
The third block lists the Kida et al. (2008) events,
and the fourth block lists the two-months 5\,GHz event detected
by Bower et al. (2007).
This last event is shown here for completeness.
References: (1) Bower et al. (2007); (2) Kida et al. (2008).
We note that the Kida et al. (2008) transients
have positional uncertainties of order
$0.4^{\circ}$ in declination,
and $0.04^{\circ}$ in right ascension.}
\label{tab:FastTran}
\end{deluxetable*}

\section{VLA\,J172059.9$+$385226.6: A re-analysis}
\label{sec:VLA1720}

VLA\,J172059.9$+$385226.6 was identified as a 9-mJy source in the FIRST survey
but was undetected in the NVSS ($S\ltorder3.5$\,mJy).
%
%
Each image in the FIRST survey
is made from data taken several days apart
(R. Becker, personal communication).
Searching the VLA archive, we have found this field
was observed three times on 1994, August 8, 13 and 14.
Re-analysis of the data
shows that this
source was present only in the last 10-s
integration out of the 2.5-minute scan taken on August 8th,
and had 370-mJy flux density.
However, a closer look at the data
showed that this event was due to
a previously unknown bug in the VLA recording system;
this bug affected the FIRST survey.
Specifically, the telescopes were repointed, but
the header information was not updated.
VLA\,J172059.9$+$385226.6 is, in fact, a genuine
source at a different sky position
($\alpha=17^{h}24^{m}00.^{s}50$, $\delta=+38^{\circ}52^{'}26.^{''}6$, J2000.0).
Therefore, VLA\,J172059.9$+$385226.6 is not a real transient source.
Unfortunately, follow up visible-light {\it Hubble Space Telescope} and near-IR Keck-II
observations
were undertaken before this realization.

We note that Bower et al. (2007) did~not find evidence
for variability in their transients,
within the 20-minute integration interval.
Therefore, the Bower et al. transients
cannot be spurious sources of the same kind
(see also a detailed discussion in Bower et al. 2007).
However, given the uncertain nature of other radio 
transients (e.g., Lorimer et al. 2007; Deneva et al. 2008),
we think that some caution is warranted.

\section{Near-IR Observations of the Bower et al. field}
\label{sec:Obs}

On UTC 2008 April 28.4 we obtained a 7500-s exposure
in $K_{{\rm s}}$-band of the
Bower et al. field, with the Hale 5.08-m telescope
at Palomar observatory (P200) equipped with the
Wide-field IR Camera (WIRC).
The field-of-view of WIRC contains all
the eight radio transients found by Bower et al.
which do~not have any visible-light counterparts.

\begin{figure*}
\centerline{\includegraphics[width=17cm]{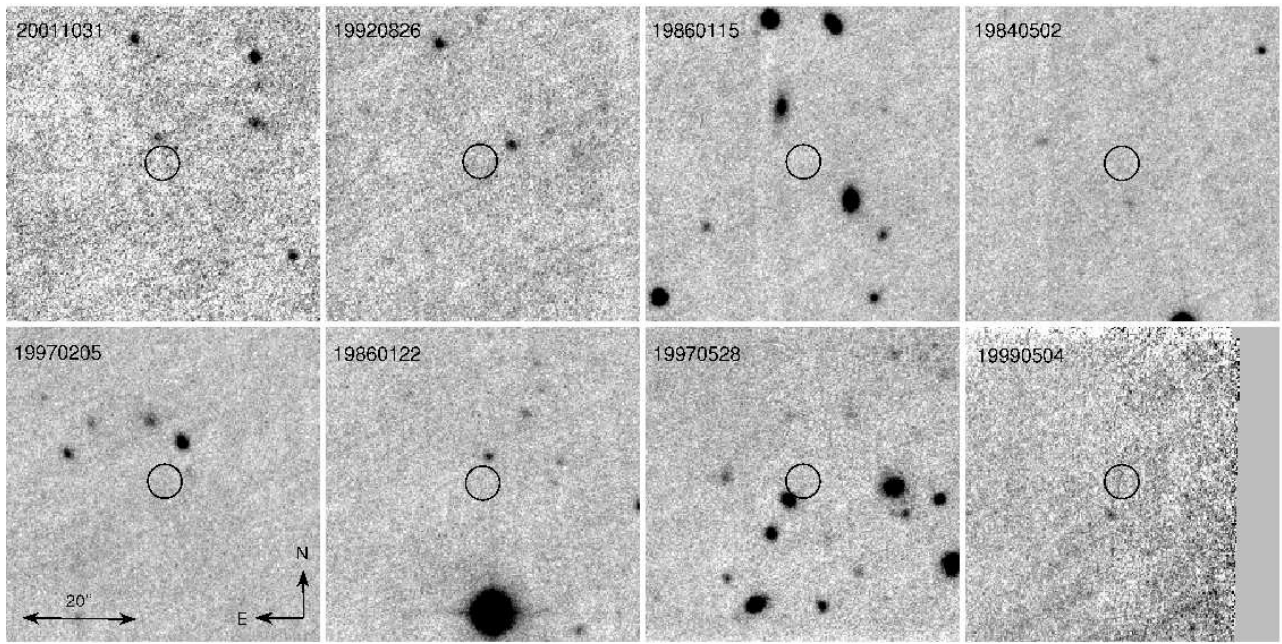}}
\caption{P200/WIRC cutouts around the locations of the eight
radio transients lacking optical counterparts found
by Bower et al. (2007).
The position of each transient is marked by a circle
with $3''$ radius.
The spatial radio position uncertainty of these transients is
about $0.3''$ (Bower et al. 2007).
The tie between the optical and radio coordinates frame
is usually better than $1''$ (e.g. Gal-Yam et al. 2006).
The transients
names are noted at top-left of each cutout
and are shown by their position East to West,
upper row from left to right and than the lower row
from left to right.
The effective exposure times for the cutouts are:
1530\,s, 3030\,s, 7500\,s, 7500\,s, 7500\,s, 5940\,s, 4440\,s,
and 1530\,s, respectively.
\label{fig:P200K_Bower_moasic}}
\end{figure*}

An astrometric solution was obtained using the \textit{ASCfit} package
(J{\o}rgensen et al. 2002) and the images were combined
using \textit{SWarp}\footnote{Written by E. Bertin; http://terapix.iap.fr/}.
Cutouts from the combined image, around the position
of the eight transients, are presented in
Figure~\ref{fig:P200K_Bower_moasic}.
We do~not detect any $K_{{\rm s}}$-band counterparts to each
of these eight transients (see Table~\ref{tab:FastTran}).

Also listed in Table~\ref{tab:FastTran}
are the flux limits, at the position
of the transients, from the {\it ROSAT}-PSPC
all-sky survey in the
0.12--2.48\,keV band (Voges et al. 1999).

\section{Observational properties of the long-duration radio transients}
\label{sec:Prop}

In the following we analyze the observational properties
of the long-duration radio transients.
Specifically, we discuss their
rate (\S\ref{sec:Rate}),
sky surface density (\S\ref{sec:SkyDen}),
and source count function (\S\ref{sec:SourceCountFun}).

\subsection{Rate}
\label{sec:Rate}

Bower et al. (2007) found that the observed areal density of
events at 5\,GHz and 8\,GHz (dominated by the 5\,GHz events),
with flux density above
370\,$\mu$Jy at a two-epoch survey, is $1.5\pm0.4$\,deg$^{-2}$.
Thus, the single epoch areal density
of events with flux greater than 370\,$\mu$Jy, is
$0.75_{-0.28,-0.45}^{+0.40,+0.81}$~deg$^{-2}$,
where the errors are given at the 1- and 2-$\sigma$ confidence
(using the formulation of Gehrels 1986).
This is translated to a 5\,GHz rate of events with flux density
above $370\,\mu$Jy of
\begin{equation}
\Re_{5\,{\rm GHz},>0.37\,{\rm mJy}} =
540_{-200,-330}^{+290,+590}
\Big( \frac{t_{{\rm dur}}}{0.5\,{\rm day}} \Big)^{-1}\,{\rm deg}^{-2}\,{\rm yr}^{-1},
\label{eq:rate5}
\end{equation}
where $t_{{\rm dur}}$ is the (unknown) typical duration of these events and
the errors are given at the 1- and 2-$\sigma$ confidence.
$t_{{\rm dur}}$ can be a function of frequency.
Therefore, comparison of this rate with rates at other frequencies should
be treated with care.
We are aware that our 5-GHz rate (Eq.~\ref{eq:rate5})
is much larger than the rate computed by Bower et al. (2007).
However, the latter estimate was the result of an arithmetic
error which once corrected yields the estimate given in Equation~\ref{eq:rate5}.

As can be gathered from Equation~\ref{eq:rate5},
the minimum rate is achieved for the largest value of $t_{{\rm dur}}$
which is seven days.
Assuming a constant event rate, this
minimum rate is
$>9\times10^{15}$ (at the $95\%$ confidence level [CL]) events over the Hubble
time.
For comparison, this estimate 
is several orders of magnitude larger than
the population of any known Galactic class of sources.
Therefore, if long-duration radio
transients are Galactic, they must be
repeaters.
On the other hand, if the events are catastrophic
(i.e., single-shot, not repeaters) then the mean
time between events is $\ltorder100$\,s (and possibly as small as $\sim 1$\,s).
The only known cosmological population with such high rate
is supernovae ($\sim1$\,s$^{-1}$; see \S\ref{sec:NPextra}).

Next we look into the rates of these events in the 1.4-GHz band.
As noted earlier (\S\ref{sec:VLA1720})
the FIRST-NVSS analysis did~not result in a firm detection
of any long-duration
transient.
The total survey area of the FIRST-NVSS search,
after correcting it for point-source incompleteness,
source confusion due to the poor resolution of the NVSS
and missing NVSS data,
is 2500\,deg$^{2}$ (see Levinson et al. 2002 for details).
We place an upper limit to the sky density
(i.e., density of sources observed in a single epoch)
of transient sources (flux density above 6\,mJy in the 1.4 GHz band)
of $1.5\times10^{-3}$\,deg$^{-2}$ and $2.6\times10^{-3}$\,deg$^{-2}$,
at the $95\%$ and $99.73\%$ CL, respectively.
We note that this sky density is consistent with
the upper limit derived by Carilli, Ivison \& Frail (2003).
Therefore, the $95\%$ confidence
upper limit on the 1.4\,GHz rate, $\Re$,
of long duration radio transients with
flux density above 6\,mJy is:
\begin{equation}
\Re_{1.4\,{\rm GHz},>6\,{\rm mJy}} <
1.1
\Big( \frac{t_{{\rm dur}}}{0.5\,{\rm day}} \Big)^{-1}\,{\rm deg}^{-2}\,{\rm yr}^{-1}.
\label{eq:rate1}
\end{equation}

\subsection{Sky Surface Density}
\label{sec:SkyDen}

Another interesting quantity is the areal density of the transients,
$\Sigma$. We obtain a lower limit by dividing the number of
unique sources by the angular area of the VLA field.  
All the Bower et al. transients were found within
$9'$ (twice the half-power radius at 5\,GHz)
from the center of a
single VLA field ($\alpha=15^{h}02^{m}20.53^{s}$, $\delta=+78^{\circ}16'14.905''$; J2000.0).
Within the half-power radius of $4.5'$
(corresponding to a solid angle of $0.018$\,deg$^{2}$)
there are four detections
(see Fig.~2 in Bower et al. 2007).
At $95\%$ CL the smallest number of sources 
is $>1.1$ (using the formulation of Gehrels 1986).
Therefore, at Galactic latitude $b\sim40^{\circ}$,
$\Sigma>60$\,deg$^{-2}$ ($95\%$ CL).

We note that the lack of any repeater events in
the Bower et al. (2007) sample can be used to improve this limit.
This can be done by calculating the probability of choosing
seven events out of $N$ with no repetition.
However, the resulting areal density is only increased by 40\% (to the
same $95\%$ CL).
We therefore retain the simpler estimate.

Assuming a constant surface density as a function
of Galactic latitude
the all-sky surface density of radio transient progenitors is
$\Sigma_{{\rm all-sky}}>3.5\times10^{6}$.
%
We note that if the long duration radio transients are
Galactic then their sky surface density toward the
Galactic plane should be higher, and therefore the
total all-sky number could be larger.

\subsection{Source number count function}
\label{sec:SourceCountFun}

The source number count function, $N(>S)$,
where $N$ is the number of events brighter than a peak flux density $S$,
may provide some hints
regarding the nature of the radio transient population.
We consider a power-law source count function,
$N(>S) \propto S^{n}$,
where
$n$ is the power-law index.
For a homogeneous population of sources residing
in an Euclidean Universe we expect $n=-3/2$,
while for Galactic thin disk population $n\approx-1$.

Assuming that $t_{{\rm dur}}$
does~not depend on the frequency,
we can use the
transient rates given in \S\ref{sec:Rate}
to put limits on the power-law index, $n$,
of the source number count function.
%
%
For each expected value of the number of events
detected in the Bower et al. (2007) survey, $\lambda_{b}$,
and the Levinson et al. (2002) search, $\lambda_{l}$,
we calculate the probability
that $k_{b}=6$ events will be detected in the Bower et al.
search and $k_{l}=0$ in the Levinson et al. survey:
\begin{equation}
P = \frac{1}{k_{l}! k_{b}!} \lambda_{b}^{k_{b}}e^{-\lambda_{b}} \lambda_{l}^{k_{l}}e^{-\lambda_{l}}.
\label{eq:PoissBL}
\end{equation}
The power-law index, $n$ relates to $\lambda_{b}/\lambda_{l}$ through:
\begin{equation}
\frac{\lambda_{b}}{\lambda_{l}}=
\frac{A_{b}}{A_{l}}
\Big(\frac{\nu_{b}}{\nu_{l}}\Big)^{n}
\Big(\frac{S_{b}}{S_{l}}\Big)^{\alpha},
\label{eq:N}
\end{equation}
where $A_{b}=9.33$\,deg$^{2}$ is the
effective\footnote{This is the area of a single epoch observation multiplied by the number of epochs.}
(single epoch) search
area of the Bower et al. survey.
This was estimated by dividing the number of transients found by Bower et al. (7),
by the one epoch search areal density (0.75\,deg$^{-2}$).
$A_{l}=2500$\,deg$^{2}$ is the area of the Levinson et al. search,
$\nu_{b}=5$\,GHz and $\nu_{l}=1.4$\,GHz are the surveys frequencies,
$S_{b}=0.37$\,mJy and $S_{l}=6$\,mJy are their specific flux limits,
and $\alpha$ is the spectral power law index of the sources ($S_{\nu}\propto \nu^{\alpha}$).
Equation~\ref{eq:N} contains three parameters: $n$, $\lambda_{l}$
and $\lambda_{b}$.
However $\lambda_{l}$ is a function of $n$ and $\lambda_{b}$.
Thus we need only explore the phase space of two
of these (e.g., $n$ and $\lambda_{b}$).

\begin{figure}
\centerline{\includegraphics[width=8.5cm]{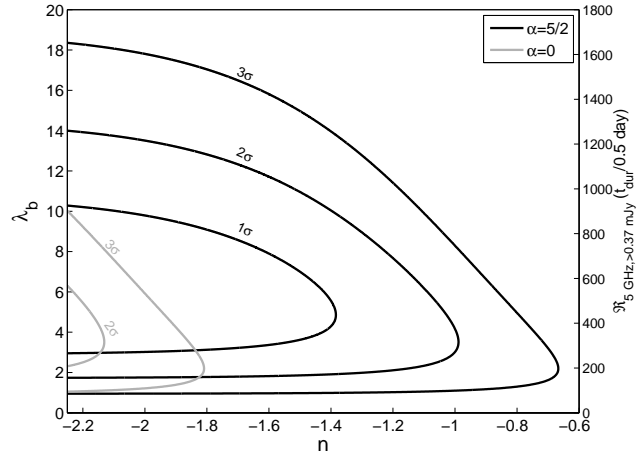}}
\caption{Confidence interval contours
(as calculated from the log-likelihood)
as a function of the expected value
of the number of events in the Bower et al. (2007) search,
$\lambda_{b}$,
and the power-law index $n$,
of the source number count function (see Eq.~\ref{eq:PoissBL}).
The {\it gray} lines show the 2- and 3-$\sigma$ contours
for spectral index $\alpha=0$,
while the {\it black} lines show the 1-, 2- and 3-$\sigma$
contours for $\alpha=5/2$.
The right-hand y-axis is the
corresponding transients rate assuming
$t_{{\rm dur}}=0.5$\,day.
\label{fig:LambdaB_n}}
\end{figure}

We calculated the probability in Equation~\ref{eq:PoissBL}
as a function of $n$ and $\lambda_{b}$.
In Figure~\ref{fig:LambdaB_n} we show the log likelihood contours
as a function of $n$ and $\lambda_{b}$ for $\alpha=5/2$
and for $\alpha=0$.
The contours are for the 1, 2 and 3-$\sigma$,
assuming two degrees of freedom (Press et al. 1992).

For $\alpha=0$, we find $n<-1.8$ at the 3-$\sigma$ CL.
Since such a steep number count index is unlikely for astrophysical
sources, we conclude that most probably $\alpha\gtorder0$.
The highest $\alpha$ possible for continuum emission is $\alpha=5/2$
(synchrotron self absorption; Rybicki \& Lightman 1979, p.~186).
For such $\alpha$ we find that $n<-0.7$ ($-1.0$) at the 3 (2)-$\sigma$
CL.

There are at least two caveats in our analysis.
First, we assumes that the
source luminosity function does~not depend on distance.
Second, the Levinson et al. (2002) and Bower et al. (2007)
rates were measured at different celestial positions.
Therefore, if the long-duration radio transients
are Galactic sources, then their sky distribution
is probably not uniform, and this may affect the results
presented in this section (however, see discussion in \S\ref{sec:IntroNS}).

\section{The progenitors of long-duration radio transients}
\label{sec:NotProgenitor}

In this section we list astrophysical sources and phenomenon
that may be responsible for the long-duration radio transients.
Some of these possibilities were already presented by Bower et al. (2007).
We first discuss the
extragalactic hypothesis (\S\ref{sec:NPextra})
followed by Galactic progenitors (\S\ref{sec:NPgal}).

\subsection{Extragalactic sources}
\label{sec:NPextra}

\begin{figure}
\centerline{\includegraphics[width=8.5cm]{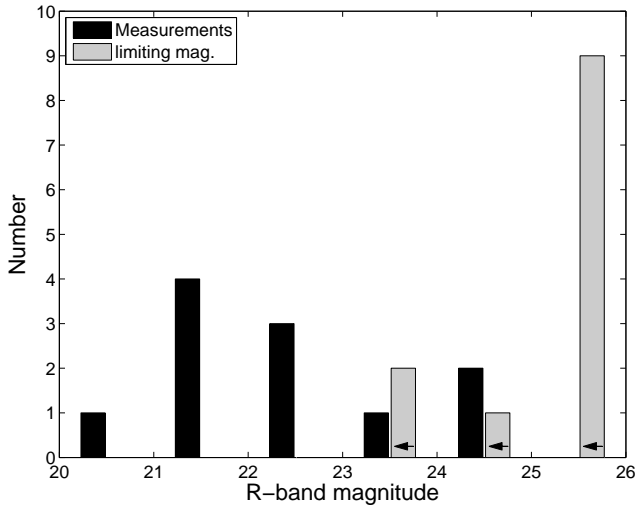}}
\caption{$R$-band magnitude ({\it black} bars) or 
$2$-$\sigma$ limiting magnitude ({\it gray} bars)
distribution of GRB host galaxies (Ovaldsen et al. 2007).
In some cases the observations were done in a different band than
$R$-band, in these cases, we converted the host galaxies
magnitude to $R$-band by assuming it is an Sc-type galaxy at
redshift of 2, based on the galaxy spectral templates of Kinney et al. (1996).
\label{fig:GRB_Hosts}}
\end{figure}

When observing the error boxes of known types
of extragalactic explosions 
(e.g., GRBs and supernovae),
we usually detect the host galaxy
(e.g., Fruchter et al. 2006; Ovaldsen et al. 2007; Perley et al. 2009).
In Figure~\ref{fig:GRB_Hosts}
we present a histogram of the $R$-band magnitude
(or limiting magnitude)
for host galaxies of a sample of {\it Swift}
(Gehrels et al. 2004) detected GRBs (Ovaldsen et al. 2007).
A high fraction ($\sim50\%$)
of the {\it Swift} GRBs are associated with galaxies brighter than about
$R\sim25$\,mag.
For optically identified supernovae (SNe) in blind surveys
the fraction is almost $100\%$.
We note that we still do~not have deep
images, and therefore constraints on the hosts,
of the new class of bright supernovae
(Barbary et al. 2009; Quimby et al. 2009).

We find it significant that only one
of the Bower et al. transients has an optical counterpart
(see \S\ref{sec:Introduction}).
It is furthermore curious that this counterpart is
a low redshift galaxy (RT\,19840613; $z\cong0.04$).
Note the absence of any intermediate redshift counterparts.

\begin{figure}
\centerline{\includegraphics[width=8.5cm]{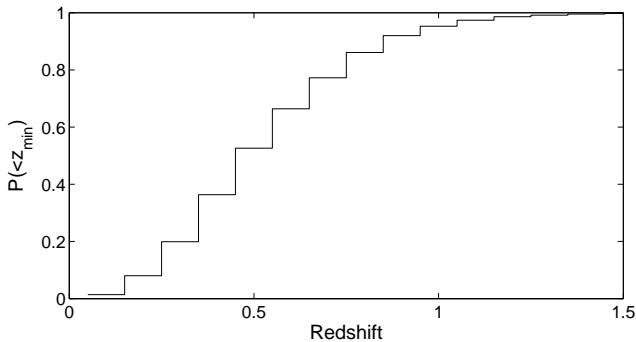}}
\caption{The cumulative probability
that the minimum redshift out of seven sources,
randomly selected from
the star formation distribution
will be smaller than a given redshift.
We used the star formation distribution
compiled by
Hopkins \& Beacom (2006) using
the parametrization of Cole et al. (2001)
and assuming a modified Salpeter IMF.
We divided this star formation distribution
by $(1+z)$ to account for the decreasing
transients rate due to time dilation.
\label{fig:P_min8z_z}}
\end{figure}

Another way to quantify this
curious absence of host galaxies
is by using the observed
star formation rate in the Universe.
In Figure~\ref{fig:P_min8z_z} we show the probability
that the minimum redshift in a sample of
seven sources randomly selected from
the observed star formation rate distribution
in the Universe (Hopkins \& Beacom 2006)
will be smaller than a given redshift.
From Figure~\ref{fig:P_min8z_z} we conclude
that at least one of the seven
radio transients will have $z<1.5$,
at the $99.73\%$ CL.
The luminosity function
of galaxies at high redshift is not well known.
However, the GRB host galaxies sample of Ovaldsen et al. (2007; Fig.~\ref{fig:GRB_Hosts})
which probe typical redshifts of $\gtorder2$
suggest that if the radio transients residing
in hosts similar to those of GRBs then
Bower et al. should have detected optical counterparts
to most of those transients.

The above discussion not withstanding we now consider the
usual suspects in the extragalactic framework.

{\bf GRBs:}
Gamma-Ray Bursts (GRBs) are often detected in radio frequencies
for periods of days to weeks (e.g., Frail et al. 1997; Chandra et al. 2008).
However, as graphically demonstrated by Figure~\ref{fig:GRB_Hosts}
the GRB hypothesis can be excluded.
Furthermore, the observed all-sky rate of GRBs is $2$~day$^{-1}$,
while the all-sky rate of long-duration radio transients is $\gtorder10^{3}$~day$^{-1}$ (\S\ref{sec:Rate}).

{\bf Orphan GRBs:}
Observational evidences suggest that GRB emission is beamed
and highly anisotropic (e.g., Harrison et al. 1999; Levinson et al. 2002).
Therefore, the actual rate of orphan GRB explosions
is $f_{b}^{-1}$ times larger than the observed rate. Here, $f_{b}^{-1}$ is the
inverse of the beaming factor,
and it is probably in the range 50--500
(e.g., Guetta, Piran \& Waxman 2005; Gal-Yam et al. 2006).
However, orphan GRB radio afterglows are expected to have time scales
of years rather than days,
which is not in line with the time scales of
long-duration radio transients.
More importantly, such orphans are expected to be at
redshifts lower than that of GRBs (which are beamed and thus seen at higher redshift)
making the host-galaxy non detections even more problematic
(see Levinson et al. 2002).



{\bf Quasars and Active Galactic Nuclei:}
To a limiting magnitude of $i\approx26$~mag,
the faintest quasars\footnote{The taxonomic definition
of quasars is nuclear absolute magnitude $M_{B}<-21.5+5\log_{10}{h_{0}}$,
where $h_{0}$ is the present day Hubble parameter
in units of 100~km~s$^{-1}$~Mpc$^{-1}$ (Peterson 1997).}
can be detected to a redshift of about 5.
The lack of optical counterparts associated with these transients
is not consistent with a quasar connection.
Low luminosity quasars (i.e., Active Galactic Nuclei; AGN),
are fainter, but if their abundance roughly follows
the star formation rate in the Universe,
then 
as shown in Figure~\ref{fig:P_min8z_z},
the host galaxies should have been found in visible light.

\begin{figure}
\centerline{\includegraphics[width=8.5cm]{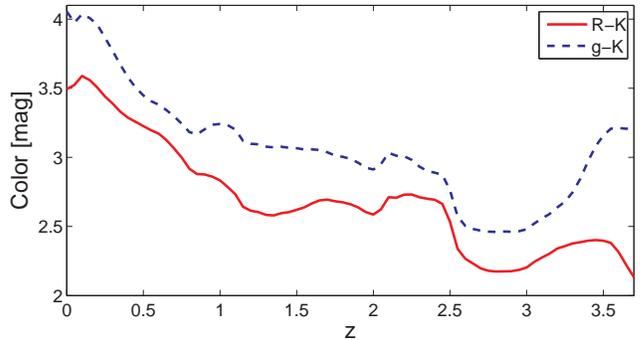}}
\caption{The $R_{{\rm Vega}}$$-$$K_{{\rm Vega}}$ ({\it solid} line),
and $g_{{\rm AB}}$$-$$K_{{\rm Vega}}$ ({\it dashed} line) colors,
of quasars, as a function of redshift $z$.
The colors are based on synthetic photometry
of quasar spectral templates (Brotherton et al. 2001; Glikman, Helfand \& White 2006).
Synthetic photometry is calculated using the code of
Poznanski et al. (2002).
The minimum $g$$-$$K$ in the range $z=0$ to $z=3.7$ is
about 2.5\,mag.
However, at $z>3.7$ the $g$$-$$K$ color will be larger since
the Lyman $\alpha$ line is found at wavelength redder
than that of the $g$-band.
\label{fig:QSO_color}}
\end{figure}

Obscured quasars, also known
as type-II quasars are faint in visible light frequencies.
However, these sources may reveal themselves in
near-IR (e.g., Gregg et al. 2002; Reyes et al. 2008).
Figure~\ref{fig:QSO_color}
shows the $g$$-$$K$ and $R$$-$$K$ color
of ``normal'' quasars as a function of redshift.
The minimum $g$$-$$K$ color of quasars
is about 2.5\,mag.
%
The non-detection of AGNs down to
$K_{{\rm s}}$-band magnitude of 20.4 (\S\ref{sec:Obs})
corresponds to a non detection
of a quasar with intrinsic $g$-band magnitude fainter than $23$.
This is based on the assumption that type-II quasars and normal quasars have similar
$K$-band luminosity functions,

To this magnitude level, the faintest quasars can be detected
up to $z\approx2$.
Therefore, the fact that we do~not
detect any near-IR sources
associated with the long-duration
radio transients
disfavors association
with reddened quasars.

{\bf Supernovae:}
There are two known variants of bright radio SNe:
type-II radio SNe, for which SN\,1979C (Weiler et al. 1991) is
the prototype, and Type-Ic radio SN
(e.g., SN\,1998bw; Kulkarni et al. 1998).

Type-II radio SNe, have long time scales (years)
and are detectable in nearby galaxies ($z\sim0.01$).
On the other hand, type-Ic radio SNe are detected
to somewhat larger distances ($z\sim0.1$),
and last a few weeks.
Therefore, based on the lack of optical counterparts
we can rule out association with type-II or type-Ic radio SNe.

{\bf Extragalactic Microlensing:}
When considering microlensing events we should discuss
the population of sources and lenses.
While the lenses can be ``unseen'' objects,
the sources should be radiant.
Most sub-mJy sources,
which are the potential sources for microlensing,
are starburst galaxies. These galaxies 
have spatial size several orders of magnitude
larger than the Einstein radius
of stellar-mass lenses.
Therefore, microlensing by extragalactic or Galactic objects
cannot explain the
high amplitude of the radio transients discussed here (Table~\ref{tab:FastTran}).

We note that the small number of background sources
and the high amplitude of the transients
also rules out scintillation events.

{\bf Extragalactic Soft Gamma-Ray Repeaters (SGRs):}
There are only eight\footnote{http://www.physics.mcgill.ca/$\sim$pulsar/magnetar/main.html}
SGRs known in the Milky Way galaxy and the Magellanic Clouds.
However, extragalactic SGRs (e.g., Eichler 2002; Nakar et al. 2006;
Popov \& Stern 2006; Ofek et al. 2006, 2008; Ofek 2007)
may be detected to larger distances.
Currently, the radio emission from giant flares
of SGRs is not well constrained (see Cameron et al. 2005).
Assuming a
fraction of $10^{-4}$ of the energy released in hard X-rays/$\gamma$-rays
is emitted in radio over one hour, a giant X-ray flare with luminosity
$L_{X}\sim10^{46}$\,erg can be observed
to a distance of a about 100\,Mpc.
However, the rate of SGR giant flares with such energy
is about $10^{-4}-10^{-5}$\,Mpc$^{-3}$\,yr$^{-1}$ (Ofek 2007).
Therefore, the expected observed rate of
Galactic and extragalactic SGRs is at least three orders
of magnitude smaller than the rate of
long-duration radio transients.


{\bf New unknown explosions:}
Explosions taking place outside galaxies
(``naked'')
may explain the lack of optical counterparts.
In general we can~not rule
out putative extra-Galactic sources, which have not been discovered yet.

We discuss several
examples of such putative sources, showing that they are unlikely
sources for the long-duration radio transients.
Hawking (1974) suggested that primordial black holes
of mass $\ltorder10^{15}$\,g will evaporate within
the Hubble time, and eventually emit a burst
of energetic photons and particles.
Such explosions are expected to manifest as a short-duration ($\ll 1$\,s) radio pulse
as the ambient magnetic field is altered by an expanding conducting shell (Rees 1977).
However, to date such events were not found
(e.g., Phinney \& Taylor 1979).
Moreover, they are expected to have very short time scales,
in contrary to long-duration radio transients.

Following the suggestion by Kulkarni et al. (2009),
Vachaspati (2008) presented a model in which
grand unification scale superconducting cosmic strings
are emitting short ($<1$\,s) radio flares.
Vachaspati (2008) predicts that the source
number count function of such events will
be $N(>S)\propto S^{-1/2}$, which is not consistent
with the long-duration radio transients source count function
(\S\ref{sec:SourceCountFun}).

Strong radio emission
from supernovae was suggested by
Colgate \& Noerdlinger (1971) and Colgate (1975).
In their scenario, the expanding core of the supernova
``combs'' the star's intrinsic dipole field,
and the generated current sheet produce
coherent radio emission.
The maximum energy emitted in such a radio pulse,
assuming no attenuation,
is of the order of $10^{46}$\,erg.
However, the expected pulse is very short ($\ll 1$\,s)
in comparison
to the observed time scale of
long-duration radio transients.
Of course, the lack of detectable host galaxies
makes any such suggestion untenable.

The lack of optical counterparts cannot rule out
association with high redshift sources
(e.g., population-III stars).
Assuming a large cosmological distance,
the rest-frame energy of such events is approximately:
\begin{eqnarray}
E_{f,cos} &\approx  &
    3\times10^{48}
    \frac{S_{\nu}}{0.4\,{\rm mJy}}
    \frac{\Delta\nu}{10^{10}\,{\rm Hz}}
    \frac{t_{{\rm dur}}}{0.5\,{\rm day}}
    \Big( \frac{d_{L}}{5\times10^{10}\,{\rm pc}} \Big)^{2} \cr
& & \times \Big( \frac{1+z}{6} \Big)
    \,{\rm erg};
\label{eq:EnPopIII}
\end{eqnarray}
%
%
here, $d_{L}$ is the luminosity distance (normalized to $1+z=6$),
$z$ is the redshift,
and $S_{\nu}$, $\Delta{\nu}$, and $t_{{\rm dur}}$ are given in the observed frame.

The rest-frame brightness temperature (normalize at $1+z=6$)
of such events is:
\begin{eqnarray}
T_{{\rm B,cos}}&=&S_{\nu,{\rm rest}} d_{lum}^{2} (c/\nu_{{\rm rest}})^{2} (2\pi k_{B} R_{{\rm rest}}^{2})^{-1} \cr
         &\approx &
                   7\times10^{16}
                   \frac{S_{\nu}}{0.4\,{\rm mJy}}
                   \Big( \frac{1+z}{6} \Big)^{-1} 
                   \Big( \frac{d_{L}}{5\times10^{10}\,{\rm pc}} \Big)^{2} \cr
         &  &      \times \Big( \frac{\nu}{5\,{\rm GHz}} \Big)^{-2}
                   \Big( \frac{R_{{\rm rest}}}{200\,{\rm au}} \Big)^{-2}
                   \,{\rm K},
\label{eq:Tb}
\end{eqnarray}
where we set the size of the emission region, $R_{{\rm rest}}$,
to $200$\,au which is
the light crossing time in the rest frame,
$c t_{{\rm dur,rest}}=c t_{{\rm dur}} (1+z)^{-1}$,
for $t_{{\rm dur}}=7$\,day.
Again, $S_{\nu}$, $\Delta{\nu}$, and $t_{{\rm dur}}$ are given in the observer frame,
while the subscript ``rest'' indicate quantities at the rest frame.
As $t_{{\rm dur}}\ltorder7$\,day, this is a lower limit
on the brightness temperature.
The huge brightness temperature,
relative to the minimum energy or equipartition
temperature (Readhead 1994), requires
a coherent emission mechanism or
alternatively a Lorentz factor of $\gtorder 10^{2}$.
However, if the beaming factor is large,
then the rate of long-duration radio transients
will exceed the SN rate in the Universe.

\subsection{Galactic}
\label{sec:NPgal}

{\bf Stellar sources:}
Several types of Galactic sources, which are known
to flare at radio wavebands were discussed by Bower et al. (2007).
Specifically, RS CVn stars, FK Com stars, Algol class binaries
and X-ray binaries are ruled out by the lack of optical counterparts.
Other possibilities, like T~Tau stars, are associated with
star forming regions typically found at low Galactic latitude.

Late-type stars are known to flare in radio wavebands
(see G\"{u}del 2002 for review).
However, the lack of $R$- and $K$-band optical/near-IR counterparts
rules out an association with late-type M-dwarfs
up to a distance of 2\,kpc, which is
the median distance of Galactic M-dwarfs at the direction
of the Bower et al. field (Gould, Bahcall, \& Flynn 1996).
Moreover, flares from M dwarfs (and brown dwarfs)
are known to exhibit
strong circular polarization
($\sim100\%$; e.g., Hallinan et al. 2007; Berger 2006).
However, as summarized in \S\ref{sec:Introduction},
Bower et al. (2007) did~not find evidence for circular
polarization above $30\%$ in any of their radio transients.

{\bf Brown dwarfs:}
In recent years, it was found that
at least some brown-dwarfs are active, and show bursts
in radio and X-ray wavebands
(e.g., Burgasser et al. 2000; Berger et al. 2001;
Berger et al. 2002; Burgasser \& Putman 2005).
In the radio regime, these flares
peaks
around 5\,GHz (spectral slopes $\alpha\gtorder0$;
Berger et al. 2001; 2005; 2008a;b).

Brown dwarfs are potentially very faint sources.
Our near-IR search of the Bower et al. (2007) field
can detect a T5-type brown dwarf 
up to distances of about 250\,pc .
However, our observations cannot rule out
an association of the Bower et al. transients
with older, or less massive, and hence cooler and fainter
brown-dwarfs (e.g., Y-class brown-dwarfs;
see Burrows et al. 1997; Baraffe et al. 2003 for evolutionary models).

For cooler brown-dwarfs which are undetectable by
our $K$-band search,
based on the minimum surface density of radio transient
progenitors derived in \S\ref{sec:SkyDen},
we can put a rough lower limit on the distance.
We assume that the local density of brown dwarfs
with effective temperature above 200\,K is
$\rho_{BD}\approx0.1$\,pc$^{-3}$ (Reid et al. 1999; Burgasser et al. 2004)
and that at small distances
the distribution can be regarded as isotropic.
Agreement between the areal density
of the radio transients (\S\ref{sec:SkyDen})
to that of old brown dwarfs is obtained by having
the brown dwarfs at a typical distance
of $d\gtorder200$\,pc.

At such distances, the energy needed to produce a single long-duration
radio transient burst will be about:
\begin{eqnarray}
E_{{\rm f,BD}} &\approx  & 8\times10^{30}
                           \frac{S_{\nu}}{0.4\,{\rm mJy}}
                           \Big( \frac{d}{200\,{\rm pc}}\Big)^{2}
                           \frac{\Delta\nu}{10^{10}\,{\rm Hz}} \cr
               &  &        \times \frac{t_{{\rm dur}}}{0.5\,{\rm day}}\,{\rm erg}.
\label{eq:EnGal}
\end{eqnarray}

Should the Bower et al. radio transients arise from
brown dwarfs then
they must repeat.
The flare repetition time scale is $\Sigma/\Re$.
For the minimum sky surface density of $\Sigma\sim60$\,deg$^{-2}$
(\S\ref{sec:SkyDen}),
and assuming a rate
$\Re\approx 500$\,yr$^{-1}$\,deg$^{-2}$,
the repetition time scale is 
$\gtorder50$\,day.

For a typical distance of 200\,pc, the total energy emitted over the
Hubble time from a single active Brown-dwarf will be $\sim10^{42}$\,erg.
For comparison, the magnetic energy of active late-type M-dwarfs
is:
\begin{eqnarray}
E_{{\rm B}} &=&
                \frac{4}{3}\pi R_{*}^{3}
                \frac{B^{2}}{8\pi} \cr
            &\cong &
                5.1\times10^{35}
                \Big( \frac{r_{*}}{70000\,{\rm km}} \Big)^{3}
                \Big( \frac{B}{3\,{\rm kG}} \Big)^{2} {\rm erg},
\label{eq:Emagnetic}
\end{eqnarray}
where $R_{*}$ is the star radius, and $B$ is its magnetic field.
We note that magnetic fields in the most magnetically active M-dwarfs
reach several kG
(e.g., Valenti, Marcy, \& Basri 1995; Johns-Krull \& Valenti 1996; Berger et al. 2008b).
This is several orders of magnitude smaller than the total
energy output of a long-duration radio transient.
Therefore, this scenario requires that the magnetic field of brown-dwarfs
will be replenished from an internal energy source.
This is not an implausible suggestion.
For example, $10^{42}$\,erg is only a small fraction
of the brown dwarf thermal or rotational energy.
The strongest argument against
the association of long-duration radio transients
with brown dwarfs is the lack of circular polarization
observed in the Bower et al. (2007) events.

{\bf Reflected solar flares:}
As discussed by Bower et al. (2007),
a solar flare reflected by a solar system object
will have a flux of:
\begin{eqnarray}
S_{{\rm ref}} &= &
                  6.6 \frac{S_{\odot,flare}}{10^{8}\,{\rm Jy}}
                  \frac{A}{0.6} \Big(\frac{d}{100\,{\rm km}}\Big)^{2} \cr
             & &  \times \Big(\frac{r}{1\,{\rm au}}\Big)^{-2}
                  \Big(\frac{\Delta}{1\,{\rm au}}\Big)^{-2}\,\mu{\rm Jy},
\label{SolarFlareAster}
\end{eqnarray}
where $S_{\odot,flare}$ is the solar flare flux density,
$r$ and $\Delta$ are the distance of the asteroid from the Sun and Earth,
respectively, and $A$ is its
albedo\footnote{The radar albedo of asteroids ranges from 0.05 for carbonaceous chondrites to 0.6 for high-density nickel-iron. Typical value is 0.15.}.
Strong X-class solar flares may have flux densities
of $10^{8}$~Jy, although they typically
last less than 20~minute (Bastian, Benz, \& Gary 1998).
%
%
Moreover, in order to explain the long-duration
radio transients using such reflected events,
we need a population of huge asteroids ($d>100$~km),
at near Earth orbits, which do~not exist.
More importantly, the motion of such objects will be detectable by the
VLA in A array whereas the Bower et al. transients are point sources.

{\bf Pulsars:}
Most of the Galactic pulsars are young objects (up to $10^{8}$\,yr),
and found near the Galactic plane.
Moreover, with the exception of magnetars, known pulsars
have radio spectrum with $\alpha\approx -2$ (e.g., Camilo et al. 2006).
Based on the catalogue of simulated NSs (Ofek 2009),
the expected surface density of NSs
with age smaller than 10 (100)\,Myr,
at the direction of the Bower et al. field,
is about 0.3 (8)\,deg$^{-2}$,
assuming $10^{9}$ NSs in the Galaxy.
We note that the number of NSs in the direction
of the Bower et al. field ($b\cong 37^{\circ}$)
rises faster than
linearly with age, since NSs are born in the Galactic
plane and ejected (mostly) to high Galactic latitudes.

{\bf Isolated old neutron stars:}
The Galaxy may host about $10^{9}$
old isolated NSs (see \S\ref{sec:IntroNS}).
This is a plausible hypothesis
and we discuss this in the next section.

To conclude, the analysis of Bower et al. (2007)
and our own analysis does not favor
association of the long-duration radio transients
with many types of astrophysical sources.
However, we cannot rule out association
of the long-duration radio transients
with Galactic isolated NSs, Galactic cool brown-dwarfs,
or some sort of exotic explosion.
Although cool brown dwarfs are interesting candidates,
they require an emission mechanism which does~not
produce circularly polarized radiation like that
observed in flare stars.

\section{Isolated-old neutron stars as progenitors of the long-duration radio transients}
\label{sec:IONS}

The organization of this section is as follows.
In \S\ref{sec:IntroNS} we 
give a brief introduction to old neutron stars.
Based on Monte-Carlo orbital simulations
of NSs we estimate their sky surface density,
distances and source number counts function (\S\ref{sec:IONS_SkyDen}).
In \S\ref{sec:E} 
we discuss the energetics
of long-duration radio transients
in the context of Galactic NS.
In \S\ref{sec:emission} we present a simple
synchrotron emission model for such radio flares,
while in \S\ref{sec:IntP} we discuss
the possibility that these are
some kind of intermittent pulsars.
Finally, we derive a lower and upper limits
on the distance scale to the Kida et al. transients
assuming they are originating from Galactic NSs (\S\ref{sec:KidaDist}).

\subsection{Introduction to Galactic isolated old NS}
\label{sec:IntroNS}

Based on the metal content of the Milky Way galaxy,
Arnett, Schramm \& Truran (1989) estimated
that
about $10^{9}$ SNe exploded in the Milky Way
and hence similar number of NSs were born in our Galaxy.
The observed SN rate in the Galaxy suggests
a number which is smaller by a factor of 3--10.
Thus it is expected that there are about $10^{8}$ to $10^{9}$
NS in the Galaxy.

NSs cool on relatively short
time scales ($\sim$$10^{6}$\,yr; e.g., Yakovlev \& Pethick 2004).
Therefore, they are expected to be intrinsically
dim and extremely hard to detect.
However, Ostriker et al. (1970) suggested that NSs
may be heated through accretion of inter-stellar matter (ISM).
This novel suggestion raised many hopes that
NSs may be detected as soft X-ray sources
(e.g., Helfand, Chanan \& Novick 1980; Treves \& Colpi 1991; Blaes \& Madau 1993).
These predictions were followed by intensive searches for such objects
(e.g., Motch et al. 1997; Maoz, Ofek, \& Shemi 1997;
Haberl, Motch, \& Pietsch 1998;
Rutledge et al. 2003;  Ag\"{u}eros et al. 2006).
However, the several candidates that were found (e.g., Haberl, Motch, \& Pietsch 1998)
are young cooling NSs
(e.g., Neuh\"{a}user \& Tr\"{u}mper 1999; Popov et al. 2000; Treves et al. 2001).

Among the usual explanations for the observational paucity of isolated
accreting old NSs
in the {\it ROSAT} all-sky survey
(Voges et al. 1999) are
the fact that the velocity distribution
of NSs is much higher than the early estimates
(e.g., Narayan \& Ostriker [1990] vs.
Cordes \& Chernoff [1998])
or the suggestion that
magnetized NSs cannot accrete
matter efficiently at the Bondi-Hoyle rate\footnote{Bondi \& Hoyle (1944).}
(e.g., Colpi et al. 1998; Perna et al. 2003).

\subsection{Surface density and distance}
\label{sec:IONS_SkyDen}

In order to estimate the surface density we need a model
for isolated old NS space distribution at the current epoch.
Ofek (2009) integrated the orbits
of simulated NSs in the Galactic gravitational potential,
using two different natal velocity
distributions and vertical scale height distributions of the
progenitor populations
suggested by
Arzoumanian et al. (2002)
and
Faucher-Gigu\`{e}re \& Kaspi (2006).
These simulations assume that $60\%$
of the Galactic NSs were born in the Galactic bulge
12\,Gyr ago, while $40\%$ were born continuously,
with a constant rate, in the disk over the past 12\,Gyr.

\begin{figure}
\centerline{\includegraphics[width=8.5cm]{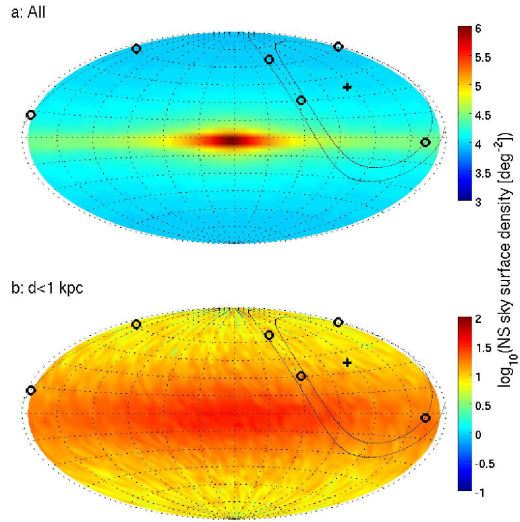}}
\caption{Sky surface density distribution of all NSs (panel a)
and NSs at distance $<1$\,kpc from the Sun (panel b).
The maps are presented in the
Galactic coordinate system and
using the Aitoff equal area projection.
The sky surface density are calculated using the catalog
of simulated old NSs presented in Ofek (2009) including both
populations of bulge-born and disk-born NSs,
and using the natal velocity distribution of
Arzoumanian et al. (2002) and assuming $10^{9}$ NSs.
The plus sign marks the position of the Bower et al. (2007) field
at $l=115^{\circ}$, $b=37^{\circ}$.
The lines mark the declination zone of 32$^{\circ}$
to 42$^{\circ}$, in which the survey for radio
transients described by Kida et al. (2008)
was conducted.
The six circles mark the position of the bright
(flux at 1.4\,GHz greater than $1$\,Jy)
transients reported by Kida et al. (2008).
\label{fig:NS_SkyDistribution}}
\end{figure}

Based on this catalog, and using the Arzoumanian et al.
initial conditions,
in Figure~\ref{fig:NS_SkyDistribution} we show the
theoretical sky surface density
of all Galactic NSs, and NSs within 1\,kpc from the Sun.
Towards the direction of the Bower et al. (2007) field,
the total surface density of isolated old NSs is about
$10^{4}$\,deg$^{-2}$.
The mean surface density of old NSs
at Galactic latitude $b>30^{\circ}$,
which roughly corresponds to the FIRST survey footprint,
is about $5\%$ larger.
The small difference between the two surface densities
suggests that if indeed long-duration
radio transients are associated with isolated old NSs
then the comparison between the
1.4\,GHz and 5\,GHz rates
presented in \S\ref{sec:SourceCountFun}
is not affected by the different sky positions at which
these surveys were conducted.

\begin{figure}
\centerline{\includegraphics[width=8.5cm]{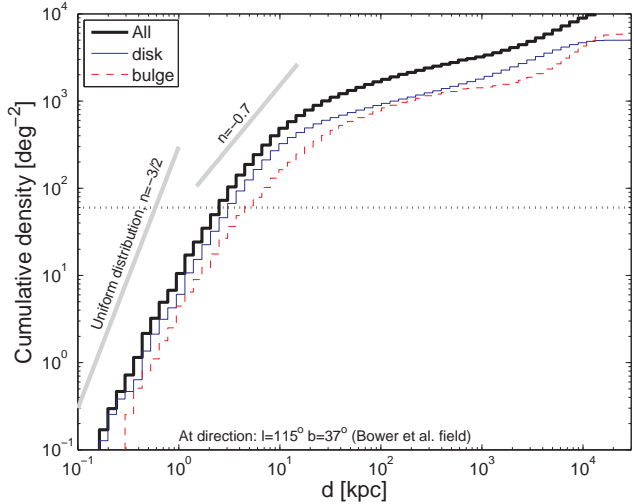}}
\caption{Cumulative number of simulated NSs per deg$^{2}$
at the direction of the Bower et al. (2007) field
as a function of distance.
The {\it thick solid} line shows the distribution of
the entire NS population
of which $60\%$ ($40\%$) are assumed to be
of bulge (disk) origin,
the {\it thin solid} line is for the disk-born NSs,
while the {\it dashed} line is for the bulge-born NSs.
The three distributions give a rough idea regarding the
uncertainties that may arise from our ignorance regarding
the birth locations of NSs along Galactic history.
The dotted horizontal line marks the minimum
surface density required ($60$\,deg$^{-2}$; see \S\ref{sec:SkyDen}).
The {\it thick solid gray} lines show the expected slopes
for a population with a homogeneous distribution ($n=-3/2$),
and $n=-0.7$, which corresponds to our $3$-$\sigma$ upper limit
(see however text).
\label{fig:NS_CumNum_BowerDirection}}
\end{figure}

In Figure~\ref{fig:NS_CumNum_BowerDirection}
we show the cumulative surface density of NSs,
at the Bower et al. (2007) field direction,
as a function of distance from the observer (i.e., the Sun).
The plots assume that the Sun is located 8\,kpc
from the Galactic center (Ghez et al. 2008).
In Figure~\ref{fig:NS_CumNum_BowerDirection}
we also mark, by a dotted horizontal line,
the minimal sky surface density
of long-duration radio transients (i.e., $\Sigma>60$\,deg$^{-2}$)
that we derived in \S\ref{sec:SkyDen}.
This Figure suggests that if isolated old NSs
are indeed related to the sub-mJy level Bower et al. transients
then their typical distance scale is at least
$\sim1$\,kpc. Otherwise the predicted sky surface density
will~not be consistent
with the minimum sky surface density of NS at the direction
of the Bower et al. field (see \S\ref{sec:SkyDen}).
On the other hand, the typical distance is presumably
not greater than about 5\,kpc, otherwise
the slope of the cumulative distribution
will be too shallow relative
to the steep
power-law index, $n$, of the number count distribution
derived in \S\ref{sec:SourceCountFun}.
We note that the luminosity function of these hypothetical
events may depend on the age (and therefore distance) to the NSs.
Therefore, we do~not attempt to quantify
the upper limit on the distance mentioned above.

A possibility that we should consider is
that only a fraction of the Galactic NSs
are the progenitors of the long-duration radio
transients (e.g., only ``young'' NSs).
Based on the NSs orbital simulations,
we find that all the NSs
with ages smaller than at least 1\,Gyr 
are required as
progenitors of the radio transients.
Thus, the sub-mJy long duration radio
transients cannot arise from pulsars.

\begin{figure}
\centerline{\includegraphics[width=8.5cm]{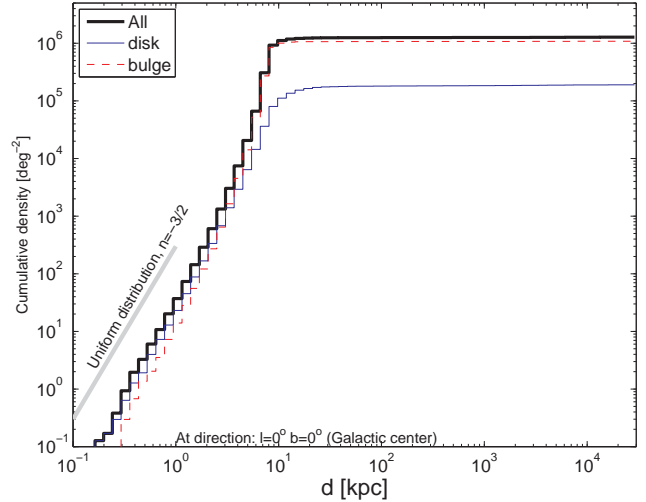}}
\caption{Like Fig.~\ref{fig:NS_CumNum_BowerDirection},
but towards the direction of the Galactic center.
\label{fig:NS_CumNum_GalCen}}
\end{figure}

A simple test for the hypothesis that long-duration
radio transients are associated with Galactic
isolated old NSs
(or for that matter any Galactic population with radial scale length of a kpc or so)
is to look
for excess of sub-mJy radio transients near the Galactic center,
relative to high Galactic latitude.
In Figure~\ref{fig:NS_CumNum_GalCen} we therefore show
the cumulative distribution of isolated old NSs as a function
of distance, but towards 
the direction of the Galactic center.

There are several marked differences between 
Figures~\ref{fig:NS_CumNum_BowerDirection} and \ref{fig:NS_CumNum_GalCen}.
The total sky surface density in the direction
of the Galactic center is about two orders of magnitude larger
than the sky surface density at the direction
of the Bower et al. field.
Next, if we consider only NSs with distances up to 1\,kpc,
then the sky surface density in the direction
of the Galactic center is about seven times larger
than that in the Bower et al. field.
Finally, the source count function in the direction
of the Galactic center is steeper than that
in the direction of the Bower et al. (2007) field.

\subsection{Energetics and time scales}
\label{sec:E}

In the old-NS framework the typical energy release from
a single flare is:
\begin{eqnarray}
E_{{\rm f}} & \approx & 2\times10^{32}
                        \frac{S_{\nu}}{0.4\,{\rm mJy}}
                        \Big( \frac{d}{1\,{\rm kpc}} \Big)^{2} \cr
            &  &        \times \frac{\Delta\nu}{10^{10}\,{\rm Hz}}
                        \frac{t_{{\rm dur}}}{0.5\,{\rm day}}\,{\rm erg}.
\label{eq:EnGal}
\end{eqnarray}
The rate of the sub-mJy long-duration radio transients is
in the range of $\Re \approx 20$\,deg$^{-2}$\,yr$^{-1}$ to $2\times10^{4}$\,deg$^{-2}$\,yr$^{-1}$ (Eq.~\ref{eq:rate5}).
Adopting a representative rate of $\Re\approx 500$\,deg$^{-2}$\,yr$^{-1}$,
and
$t_{{\rm dur}}\sim0.5$\,day we find that
over the Hubble time the all-sky number of events is
larger by several orders of magnitude than that
of any known Galactic stellar population.
Therefore, any hypothesis involving Galactic stars
would require that the sources be repeaters.
The mean time scale between bursts, $\Delta{t_{{\rm b}}}$,
is given by:
\begin{eqnarray}
\Delta{t_{{\rm b}}} &= &\frac{\Sigma}{\Re} \cr
                    &\cong &
                             0.2\frac{\Sigma}{10^{2}\,{\rm deg}^{-2}}
                             \Big( \frac{\Re}{500} \Big)^{-1}\,{\rm yr} \cr
                    &\cong &
                             0.2\frac{\Sigma}{10^{2}\,{\rm deg}^{-2}}
                             \frac{t_{{\rm dur}}}{0.5\,{\rm day}} \,{\rm yr},
\label{eq:RepT}
\end{eqnarray}
where $\Sigma$ is the sky surface density of sources in the
direction of the Bower et al. (2007) field.
In that case the flares duty cycle is of order of
$\mathcal{D}=t_{{\rm dur}}/\Delta{t_{b}}\approx7\times10^{-3}$.
An immediate prediction is that if
long-duration radio transients are associated with
Galactic NSs, then the repetition time scale for
bursts is relatively short, of the order of several months.

Multiplying Equation~\ref{eq:EnGal} by the expected number of flares
within the Hubble time
($t_{{\rm H}}/\Delta{t}_{{\rm b}} \sim 5\times10^{10}$, where $t_{{\rm H}}$ is the Hubble time),
we find that the total energy emitted by a single object
over its life time is:
\begin{eqnarray}
E_{{\rm tot}} &\approx & 1\times 10^{43}
                         \frac{S_{\nu}}{0.4\,{\rm mJy}}
                         \Big( \frac{d}{1\,{\rm kpc}} \Big)^{2}
                         \frac{\Delta\nu}{10^{10}\,{\rm Hz}} \cr
              &        & \times
                         \Big( \frac{\Sigma}{10^{2}\,{\rm deg}^{-2}} \Big)^{-1}
                         \,{\rm erg},
\label{eq:TotE}
\end{eqnarray}

Dividing Equation~\ref{eq:EnGal} by Equation~\ref{eq:RepT},
the mean luminosity over time required in order to power these bursts is:
\begin{eqnarray}
\langle\dot{E}\rangle &\approx & 3\times10^{25}  
                                 \frac{S_{\nu}}{0.4{\rm mJy}}
                                 \Big( \frac{d}{1\,{\rm kpc}} \Big)^{2}
                                 \frac{\Delta\nu}{10^{10}\,{\rm Hz}} \cr
                      &        & \times 
                                 \Big( \frac{\Sigma}{10^{2}\,{\rm deg}^{-2}} \Big)^{-1}
                                 \,{\rm erg\,s}^{-1}.
\label{eq:RepMeanE}
\end{eqnarray}
Note that this quantity is independent of $\Delta t_{{\rm b}}$.

Next, we consider the energy reservoir of isolated old NSs,
and check whether it is consistent with
the mean luminosity given by
Equation~\ref{eq:RepMeanE}.
Isolated old NSs have several sources of available energy.
Among these are:
(i) spin energy;
(ii) magnetic energy;
and (iii) accretion, from the ISM, energy.

The rotational kinetic energy of NSs is:
\begin{eqnarray}
E_{{\rm rot}} & =     & \frac{1}{2}I\omega^{2} \cr
              & \cong & 2.2\times10^{44}
                        \Big( \frac{P}{10\,{\rm s}} \Big)^{-2}
                        \frac{M_{{\rm NS}}}{1.4\,{\rm M}_{\odot}}
                        \Big( \frac{R_{{\rm NS}}}{10\,{\rm km}} \Big)^{2}\,{\rm erg},
\label{eq:Erot}
\end{eqnarray}
where $M_{{\rm NS}}$ is the NS mass, $I$ is the moment of inertia (assuming $I=0.4M_{{\rm NS}}R_{{\rm NS}}^{2}$),
$\omega$ is the NS rotational angular speed, and $P$ ($=2\pi/\omega$)
is its rotation period.
The energy-loss rate available from the rotational energy
reservoir is:
\begin{eqnarray}
\dot{E}_{\rm rot} & = & I \omega \dot{\omega} = 4\pi^{2} I P^{-3} \dot{P} \cr
                  & \cong & 4.4\times10^{26}
                            \frac{M_{{\rm NS}}}{1.4\,{\rm M}_{\odot}}
                            \Big( \frac{R_{{\rm NS}}}{10\,{\rm km}} \Big)^{2} \cr
                  &       & \times \Big( \frac{P}{10\,{\rm s}} \Big)^{-3}
                            \frac{\dot{P}}{10^{-17}}\,{\rm erg\,s}^{-1},
\label{eq:Lumrot}
\end{eqnarray}
where $\dot{\omega}$ is the time derivative of the NS angular frequency,
and $\dot{P}$
is defined as minus of its period derivative.
Typical values for $\dot{P}$ are in the range
of $10^{-10}$\,s\,\.{s} for magnetars,
$10^{-17}$\,s\,\.{s} to $10^{-12}$\,s\,\.{s} for normal radio pulsars,
and 
$10^{-21}$\,s\,\.{s} to $10^{-18}$\,s\,\.{s} for millisecond pulsars
(Srinivasan 1989).

A large fraction of NSs may have high surface magnetic field
in excess of $10^{14}$\,G (e.g., Magnetars).
The energy stored in the magnetic field of such objects is about:
\begin{eqnarray}
E_{{\rm B}} & \approx & \frac{4}{3}\pi R_{{\rm NS}}^{3} \frac{B^{2}}{8\pi} \cr
            & \approx & 1.7\times10^{45} \Big( \frac{r_{{\rm NS}}}{10\,{\rm km}} \Big)^{3}
                        \Big( \frac{B}{10^{14}\,{\rm G}} \Big)^{2} {\rm erg},
\label{eq:Emagnetic}
\end{eqnarray}
where $R_{{\rm NS}}$ is the NS radius, and $B$ is its interior mean magnetic field.
We note that
the internal magnetic field may
be higher, and therefore, Equation~\ref{eq:Emagnetic}
is a lower limit on the magnetic energy reservoir.
The rate at which the magnetic field of NSs is decaying is
debated both theoretically
and observationally
(e.g., Urpin \& Muslimov 1992; Chanmugam 1992; Phinney \& Kulkarni 1994; Sengupta 1997; Sun \& Han 2002).
Assuming the magnetic field is uniformly
decaying on the Hubble time scale,
this will provide, on average, energy loss rate
of $\dot{E}\approx4\times10^{27}$\,erg\,s$^{-1}$.

Another possible source of energy is heating of the NS by accretion from
the ISM (e.g., Ostriker et al. 1970).
Assuming NSs accrete at the Bondi-Hoyle rate (Bondi \& Hoyle 1944),
the energy-loss rate will be:
\begin{eqnarray}
\dot{E}_{{\rm acc}} & =      & \frac{4\pi G^{3} M_{{\rm NS}}^{3} \rho_{{\rm ISM}}}{R_{{\rm NS}} (v^{2} + c_{s}^{2})^{3/2} } \cr
                    &\approx & 5\times10^{26}
                               \Big( \frac{M_{{\rm NS}}}{1.4\,{\rm M}_{\odot}} \Big)^{3}
                               \Big( \frac{R_{{\rm NS}}}{10\,{\rm km}} \Big)^{-1} \cr
                    &        & \times \frac{n_{{\rm ISM}}}{0.1 {\rm cm}^{-3}}
                               \Big( \frac{v}{300\,{\rm km\,s}^{-1}} \Big)^{-3}\,{\rm erg}\,{\rm s}^{-1},
\label{eq:Eacc}
\end{eqnarray}
where $v$ is the velocity of the NS relative to an
ISM with a mass (number) density $\rho_{{\rm ISM}}$ ($n_{{\rm ISM}}$).
$c_{s}$ is the sound speed in the ISM,
which is of the order of 10\,km\,s$^{-1}$
and is therefore
neglected\footnote{The sound speed is given by $c_{s}=(\gamma k_{B}T/\mu_{m})^{1/2}$,
where $\gamma$ is the adiabatic index, $k_{B}$ is the Boltzmann constant, $T$ is the temperature,
and $\mu_{m}$ is the mean weight of the ISM particles.
For $\gamma=5/3$ and $T=10^{4}$\,K the sound speed is 11 and 15\,km\,s$^{-1}$
for neutral and ionized gas, respectively.}.
We note, however, that magneto hydrodynamic simulations suggest that in the presence of a strong
magnetic field the accretion rate will be suppressed
relative to the Bondi rate
(e.g., Toropina et al. 2001, 2003, 2005; see however Arons \& Lea 1976, 1980).

To conclude, 
the combination of NSs energy sources
may provide enough energy to explain
long-duration radio transients.
Specifically, the rotation energy of NSs
and the energy available for NSs from accretion
from the ISM 
is at least an order of magnitude larger
than needed for generating the long-duration radio transients
(Eq.~\ref{eq:RepMeanE}).
However, unless additional energy sources are invoked,
this implies that the outbursts can not emit
much higher energy at other frequencies.

\subsection{Incoherent synchrotron radiation from an afterglow}
\label{sec:emission}

Now we address the mechanism of radio emission.
There are two possibilities:
(i) incoherent synchrotron emission (the afterglow model);
and (ii) coherent emission.
We discuss the first possibility in this section
and the second possibility in \S\ref{sec:IntP}.

In the afterglow model, the source undergoes an explosive event and
ejects relativistic particles and may generate magnetic field
(hereafter relativistic plasma).
Some sort of pressure confinement is needed to prevent rapid expansion
of the relativistic plasma.
Otherwise the expansion or adiabatic losses vastly increase the
energy budget which would be inconsistent with the old NS framework.
In the case of GRBs, the afterglow is confined by the dynamic
pressure of the blast wave.
We return to this critical issue towards
the end of the subsection.
We proceed by computing the (quasi)static properties
of the (approximately) confined relativistic plasma.
We note that the parameters derived from
this model are estimated to within an order of magnitude.

Our simple model involves four free
parameters:
the radius of the emitting region, $R$;
the mean electron density, $n_{{\rm e}}$;
the magnetic field, $B$;
and the characteristic electron Lorentz factor, $\gamma_{{\rm e}}=(1-\beta_{{\rm e}}^{2})^{-1/2}$,
where $\beta_{{\rm e}}$ is the typical electron speed in units of the speed of light.
The large value of the 5\,GHz rate (Eq.~\ref{eq:rate5})
relative to that at 1.4\,GHz rate (Eq.~\ref{eq:rate1})
suggests that the synchrotron self absorption
frequency is above 5\,GHz (see \S\ref{sec:SourceCountFun}).
Therefore, we assume $\nu_{s}\gtorder5$\,GHz,
and that the optical depth at 5\,GHz, $\tau_{5}>1$.
Furthermore, we assume that the optical depth
at the synchrotron frequency $\tau_{s}\ltorder 1$.

In order to estimate these parameters,
we use the following relations:

(i) The characteristic synchrotron frequency:
\begin{eqnarray}
\nu_{s}  & =     & \gamma_{{\rm e}}^{2} \frac{eB}{2\pi m_{{\rm e}}c} \cr
         & \cong & 2.8\times10^{6} \gamma_{{\rm e}}^{2} B\,{\rm Hz},
\label{eq:Sync_F}
\end{eqnarray}
where $e$ is the elementary (electron) charge,
$m_{{\rm e}}$ is the electron mass, and $B$ the magnetic field
in cgs units (i.e., Gauss).

(ii) The power emitted by a relativistic single electron
due to synchrotron radiation
(e.g., Rybicki \& Lightman 1979) is:
\begin{eqnarray}
P_{s}  & =     & \frac{4}{3}\sigma_{T}c\beta^{2}\gamma_{{\rm e}}^{2}U_{B} \cr
       & \cong & 1.1\times10^{-15}
                 \beta^{2}\gamma_{{\rm e}}^{2} B^{2}\,{\rm erg}\,{\rm s}^{-1},
\label{eq:Sync_Pow}
\end{eqnarray}
where $\sigma_{T}$ is the Thomson cross-section,
and $U_{B}=B^{2}/(8\pi)$ is the magnetic field energy density.
The synchrotron cooling time scale, assuming $\tau_{s}\ltorder 1$,
is given by
\begin{eqnarray}
t_{{\rm syn}} & =     & \frac{\gamma_{{\rm e}} m_{{\rm e}}c^{2}}{P_{{\rm s}}} \cr
        & \cong & 7.7\times 10^{8}
                  \gamma_{{\rm e}}^{-1}
                  \beta^{-2}
                  B^{-2}\,{\rm s}.
\label{eq:Sync_cooling}
\end{eqnarray}

(iii) The brightness temperature, $T_{{\rm B}}$
(essentially a conveniently chosen surrogate for distance) is:
\begin{eqnarray}
T_{{\rm B}} & =      & S_{\nu} d^{2} (c/\nu)^{2} (2\pi k_{{\rm B}} R^{2})^{-1} \cr
      & \cong  & 1.6\times10^{21}
                 \frac{S_{\nu}}{0.4\,{\rm mJy}}
                 \Big( \frac{d}{1\,{\rm kpc}} \Big)^{2} \cr
      &        & \times \Big( \frac{\nu}{5\,{\rm GHz}} \Big)^{-2}
                 \Big( \frac{R}{10\,{\rm km}} \Big)^{-2}\,{\rm K},
\label{eq:Tb_NS}
\end{eqnarray}
where $\nu$ is the frequency at which we observe (i.e., 5\,GHz).
For optical depth larger than unity the brightness temperature
is related to the electrons energy:
\begin{equation}
k_{{\rm B}} T_{{\rm B}} \approx \gamma_{{\rm e}} m_{{\rm e}} c^{2}.
\label{eq:F_nu}
\end{equation}

(iv) Finally, in order to get a self absorption spectrum
(see \S\ref{sec:SourceCountFun})
the optical depth at 5\,GHz
should be larger than unity.
This holds if and only if at 5\,GHz the thermal emission
is smaller than the optically thin emission:
\begin{equation}
2\pi k_{{\rm B}}T_{{\rm B}} \Big( \frac{\nu}{c} \Big)^{2} <
   \frac{1}{3} R n_{{\rm e}} \frac{P_{{\rm s}}}{\nu_{{\rm s}}} \Big( \frac{\nu}{\nu_{{\rm s}}} \Big)^{1/3}
\label{eq:tau_s}
\end{equation}

Next,
we can 
solve equations~\ref{eq:Sync_F}--\ref{eq:tau_s}
for the free parameters $B$, $\gamma_{{\rm e}}$, and $R$
and we can put a lower limit on the value of $n_{{\rm e}}$.
%
Since we do~not know the exact value
of $t_{{\rm dur}}$, $\nu_{{\rm s}}$ and $d$
we solve for the free parameters as a function
of these arguments.
For completeness, we also state the dependency on $S_{\nu}$.
We normalized the solutions for
$S_{\nu} = 0.4$\,mJy, $d=1$\,kpc,
$t_{{\rm syn}}\approx t_{{\rm dur}}=0.5$\,day,
and $\nu_{{\rm s}}=2\times 10^{11}$\,Hz
(the choice for $\nu_{{\rm s}}$ is to minimize the total energy; see below).
We note that $t_{{\rm dur}}\approx t_{{\rm syn}}$
means that there is no
energy injection
into the emission region after the initial burst
(i.e., the duration
of the events is dominated by the synchrotron cooling time scale).
However, if energy is injected then $t_{{\rm dur}}\gtorder t_{{\rm syn}}$.

The following solution holds for $\tau_{5}\gtorder 1$ and $\tau_{{\rm s}}\ltorder 1$:
\begin{equation}
B\approx
17
\Big( \frac{\nu_{{\rm s}}}{2\times 10^{11}\,{\rm Hz}} \Big)^{-1/3}
\Big( \frac{t_{{\rm syn}}}{0.5\,{\rm day}} \Big)^{-2/3}
\,{\rm G},
\label{Eq:syn_B}
\end{equation}
\begin{equation}
\gamma_{{\rm e}}\approx
66
\Big( \frac{\nu_{{\rm s}}}{2\times10^{11}\,{\rm Hz}} \Big)^{2/3}
\Big( \frac{t_{{\rm syn}}}{0.5\,{\rm day}} \Big)^{1/3},
\label{Eq:syn_gamma}
\end{equation}
\begin{eqnarray}
R & \approx & 6\times10^{10}
              \Big( \frac{\nu_{{\rm s}}}{2\times10^{11}\,{\rm Hz}} \Big)^{-1/3}
              \Big( \frac{t_{{\rm syn}}}{0.5\,{\rm day}} \Big)^{-1/6} \cr
  &         & \times
              \Big( \frac{S_{\nu}}{0.4\,{\rm mJy}} \Big)^{1/2}
              \Big( \frac{d}{1\,{\rm kpc}} \Big)
              \,{\rm cm},
\label{Eq:syn_R}
\end{eqnarray}
and the lower limit on $R n_{{\rm e}}$ is

\begin{equation}
R n_{{\rm e}} > 2\times10^{16} \Big( \frac{\nu_{{\rm s}}}{2\times10^{11}\,{\rm Hz}} \Big)^{4/3}
                          \frac{t_{{\rm syn}}}{0.5\,{\rm day}} \,{\rm cm}^{-2}.
\label{Eq:sym_Ne}
\end{equation}
We note that in order to minimize the energy, $n_{{\rm e}}$ should be 
around the lower limit implied by Equation~\ref{Eq:sym_Ne}.
For convenient we also give the brightness temperature:
\begin{eqnarray}
T_{{\rm B}} & \approx & 4\times10^{11}
                  \Big( \frac{\nu_{{\rm s}}}{2\times 10^{11}\,{\rm Hz}} \Big)^{2/3} \cr
      &         & \times
                  \Big( \frac{t_{{\rm syn}}}{0.5\,{\rm day}} \Big)^{1/3} 
                  \,{\rm K}.
\label{Eq:sym_Tb}
\end{eqnarray}

Next, we can derive the ratio between
the magnetic energy and electron energy densities:
\begin{eqnarray}
U_{{\rm B}}/U_{{\rm e}} & =       & \frac{B^{2}/(8\pi)}{n_{{\rm e}}\gamma_{{\rm e}}m_{{\rm e}}c^{2}} \cr
            & \ltorder  & 1
                        \Big( \frac{\nu_{{\rm s}}}{2\times 10^{11}\,{\rm Hz}} \Big)^{-3}
                        \Big( \frac{t_{{\rm syn}}}{0.5\,{\rm day}} \Big)^{-17/6} \cr
            &         & \times
                        \Big( \frac{S_{\nu}}{0.4\,{\rm mJy}} \Big)^{1/2}
                        \Big( \frac{d}{1\,{\rm kpc}} \Big),
\label{eq:UBUe}
\end{eqnarray}
where $U_{{\rm e}}=n_{{\rm e}}\gamma_{{\rm e}} m_{{\rm e}}c^{2}$
is the electrons energy density.
We note that $U_{{\rm B}}/U_{{\rm e}}$
is very sensitive to both the duration, $t_{{\rm syn}}$,
and the synchrotron frequency, $\nu_{{\rm s}}$,
whose values are not well known.
Therefore, a small change in these unknown parameters
will change $U_{{\rm B}}/U_{{\rm e}}$ dramatically (see also Readhead 1994).

We note that by setting $U_{{\rm B}}/U_{{\rm e}}\sim1$,
we have ensured that $T_{{\rm B}}$ is equal to
the equipartition brightness
temperature (Readhead 1994) and minimized the energy requirement.
Next we test if these parameters
are below the inverse-Compton catastrophe limit
(Kellermann \& Pauliny-Toth 1969).
The inverse-Compton catastrophe is relevant
if $U_{{\rm rad}}/U_{{\rm B}}\gtorder1$.
Assuming $\tau_{{\rm s}}\ltorder 1$:
\begin{equation}
U_{{\rm rad}} \approx \frac{R U_{{\rm e}}}{c t_{{\rm syn}}}.
\label{eq:Urad}
\end{equation}
Since $R/(c t_{{\rm syn}})\ll 1$
we get $U_{{\rm rad}}\ll U_{{\rm e}}\sim U_{{\rm B}}$.
Therefore, inverse-Compton effects can be neglected.

\begin{figure}
\centerline{\includegraphics[width=8.5cm]{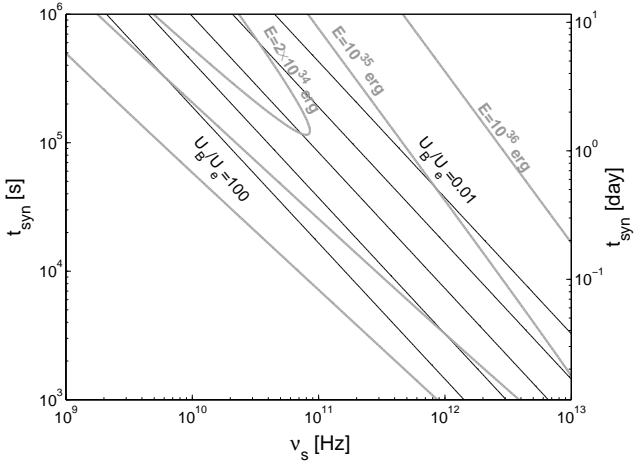}}
\caption{$U_{{\rm B}}/U_{{\rm e}}$ ({\it black} contours) and total energy ({\it gray} contours), $E=4/3\pi R^{3} (U_{{\rm e}} + U_{{\rm B}})$, as a function of $\nu_{{\rm s}}$ and $t_{{\rm syn}}$, and assuming $d=1$\,kpc, and $S_{\nu}=0.4$\,mJy.
The equal $U_{{\rm B}}/U_{{\rm e}}$ lines are 100, 10, 1, 0.1, and 0.01
from left to right.
\label{fig:syn_model_Nu_s_T_dur}}
\end{figure}

Assuming the electron density is close to the minimum density
implied by Equation~\ref{Eq:sym_Ne},
the total energy in a single burst
(i.e., in both the electrons and magnetic field)
is given by:
\begin{eqnarray}
E & =     & \frac{4}{3}\pi R^{3} (U_{{\rm e}} + U_{{\rm B}}) \cr
  & \cong & 1\times10^{34}
            \Big[
            \Big( \frac{\nu_{{\rm s}}}{2\times10^{11}\,{\rm Hz}} \Big)^{4/3}
            \Big( \frac{t_{{\rm syn}}}{0.5\,{\rm day}} \Big) \cr
  &       & \times
            \Big( \frac{S_{\nu}}{0.4\,{\rm mJy}} \Big)
            \Big( \frac{d}{1\,{\rm kpc}} \Big)^{2}
            \Big] \cr
  &       & + 1\times10^{34}
            \Big[
            \Big( \frac{\nu_{{\rm s}}}{2\times10^{11}\,{\rm Hz}} \Big)^{-5/3}
            \Big( \frac{t_{{\rm syn}}}{0.5\,{\rm day}} \Big)^{-11/6} \cr
  &       & \times 
            \Big( \frac{S_{\nu}}{0.4\,{\rm mJy}} \Big)^{3/2}
            \Big( \frac{d}{1\,{\rm kpc}} \Big)^{3}
            \Big]\,{\rm erg}.
\label{Eq:syn_TotalPower}
\end{eqnarray}
The specific values we have selected,
$\nu_{{\rm s}}=2\times 10^{11}$\,Hz
and $t_{{\rm syn}}=0.5$\,day,
give a solution which is near equipartition
and therefore minimizes the energy.
For $n_{{\rm e}}$ close to its minimum implied by
Equation~\ref{Eq:sym_Ne},
the total energy in a burst,
and $U_{{\rm B}}/U_{{\rm e}}$ as a function
of $\nu_{{\rm s}}$ and $t_{{\rm syn}}$ are shown in Figure~\ref{fig:syn_model_Nu_s_T_dur}.
This Figure suggests that the minimum energy required per burst
is around $10^{34}$\,erg.
This energy per burst 
multiplied by the expected number of bursts over the Hubble time ($\sim 5\times10^{10}$)
is of the same order of magnitude of the
energy reservoir of NSs identified in \S\ref{sec:E}.

So far we have assumed that the duration of the events
is set by the synchrotron cooling timescale.
We now explore the consequences of decreasing the cooling
timescale and letting the duration be determined by
the plasma injection timescale.
In this case Equation~\ref{Eq:syn_TotalPower} represent the total energy
of a flare within the synchrotron cooling time scale.
Therefore, in order to get the total energy we need
to multiply Equation~\ref{Eq:syn_TotalPower} by $t_{{\rm dur}}/t_{{\rm syn}}$.
As can be gathered from Figure~\ref{fig:syn_model_Nu_s_T_dur}
the energy is minimized by
setting the cooling time equal to
the duration time.

It is interesting to compare the derived radius,
$R\approx 6\times10^{10}$\,cm, with some
typical radii dominating NS physical processes.
The light-cylinder radius is:
\begin{equation}
R_{lc}=\frac{c}{2\pi}P
\cong
4.8\times10^{10} \frac{P}{10\,{\rm s}}\,{\rm cm}.
\label{Eq:R_lc}
\end{equation}
The co-rotation radius of a NS is:
\begin{eqnarray}
R_{cor} & =     & \Big( \frac{G M_{{\rm NS}}}{\omega^{2}} \Big)^{1/3} \cr
        & \cong & 7.8\times10^{8}
                  \Big( \frac{M_{{\rm NS}}}{1.4\,{\rm M}_{\odot}} \Big)^{1/3}
                  \Big( \frac{P}{10\,{\rm s}} \Big)^{2/3}\,{\rm cm}.
\label{eq:R_cor}
\end{eqnarray}
%
Assuming a Bondi-Hoyle accretion
rate:
\begin{equation}
\dot{M}= \frac{4\pi G^{2} M_{{\rm NS}}^{2} m_{p} n_{ISM}}{ (V^{2} + c_{{\rm s}}^{2})^{3/2} },
\label{BH_accRate}
\end{equation}
where $m_{p}$ is the proton mass,
then the Magnetosphere radius is:
\begin{eqnarray}
R_{m} & =      & \Big( \frac{\mu^{2}}{\dot{M}\sqrt{2 G M_{{\rm NS}}}} \Big)^{2/7} \cr
      & \cong  & 2.1\times10^{11}
                 \Big( \frac{\mu}{4\times10^{30}} \Big)^{4/7}
                 \Big( \frac{\dot{M}}{2\times10^{8}\,{\rm gr}\,{\rm s}^{-1}} \Big)^{-2/7} \cr
      &        & \times \Big( \frac{M_{{\rm NS}}}{1.4\,{\rm M}_{\odot}} \Big)^{-1/7}\,{\rm cm},
\label{Eq:Rm}
\end{eqnarray}
where $\mu$ is the magnetic dipole moment of the NS.
Finally, the accretion radius is:
\begin{eqnarray}
R_{{\rm acc}} & \approx & \frac{2 G M_{{\rm NS}}}{V^{2}} \cr
        & \approx & 1.7\times10^{12}
                    \frac{M_{{\rm NS}}}{1.4\,{\rm M}_{\odot}}
                    \Big( \frac{V}{150\,{\rm km}\,{\rm s}^{-1}} \Big)^{-2}\,{\rm cm}.
\label{Eq:Racc}
\end{eqnarray}

As noted at the beginning of the sub-section rapid
expansion of the radiating plasma would vastly increase
the energy budget.
For this reason, an integral requirement of the incoherent
model is that the plasma must be confined
(in which case the duration of the event is set by
the cooling time or by the duration of the injection
of energy by the source).
We note that the confinement radius should probably
be smaller than the light cylinder radius.
Otherwise, the energy requirement will be larger
due to the inertia of the electrons.
The equipartition radius we find
(Eq.~\ref{Eq:syn_R}) is a little bit larger than the
plausible confinement radius (e.g., Eq.~\ref{Eq:R_lc}).
However, our calculation provides only
an order of a magnitude estimate
to the fireball parameters and the equipartition radius
in particular.
Therefore, we cannot rule
out the incoherent synchrotron model
based on the small inconsistency
between the equipartition radius
and light cylinder radius.

Assuming a dipole magnetic field, decaying as $R^{-3}$
(inside the light cylinder radius; Eq.~\ref{Eq:R_lc}),
and $B=17$\,G at $R=6\times10^{10}$\,cm,
the extrapolated magnetic field strength on the surface of
a 10\,km radius NS will be about:
\begin{eqnarray}
B(10\,{\rm km}) & \approx  & B
                             \Big( \frac{R}{10\,{\rm km}} \Big)^{3} \cr
                & \approx  & 4\times 10^{15}
                             \Big( \frac{\nu_{{\rm s}}}{2\times 10^{11}\,{\rm Hz}} \Big)^{-4/3}
                             \Big( \frac{t_{{\rm syn}}}{0.5\,{\rm day}} \Big)^{-7/6} \cr
                &          & \times 
                             \Big( \frac{S_{\nu}}{0.4\,{\rm mJy}} \Big)^{3/2}
                             \Big( \frac{d}{1\,{\rm kpc}} \Big)^{3}
                             \,{\rm G}.
\label{Eq:Bsurf}
\end{eqnarray}
This is higher than the typical estimated surface magnetic
fields of pulsars and Magnetars.
However, for somewhat larger $\nu_{{\rm s}}$
or $t_{{\rm syn}}$ the discrepancy is smaller.
Moreover, the events may be related to the release of magnetic
energy stored in the NS interior.
Therefore, we conclude that the incoherent synchrotron model
cannot be ruled out.

Finally, we note that the total mass of matter within
the emission radius
is $4/3 \pi n_{{\rm e}} m_{p} R^{3}\sim 4\times10^{14}$\,gr.
Interestingly, this mass
is similar to the total amount of matter that
a NS with a space velocity of 150\,km\,s$^{-1}$
will accrete from the ISM (assuming $n=0.1$\,cm$^{-3}$ and
Bondi-Hoyle accretion)
within several months, which is the typical time interval
between bursts that we found in \S\ref{sec:E}.
We note that Treves, Colpi \& Lipunov (1993)
suggested that accreted matter from the ISM is piled
up near the Alfv\'{e}n radius, followed by
infall of the piled up matter on the NS.
They estimated that these episodic infalls may occur
every several months.

\subsection{Flares from intermittent pulsars}
\label{sec:IntP}

In recent years, several types of pulsars with
small duty cycles have been discovered.
This includes the RRATs (McLaughlin et al. 2006),
and intermittent pulsars (Kramer et al. 2006).
Several models were suggested to explain
such episodic pulsars (e.g., Treves et al 1993; Zhang, Gil \& Dyks 2006;
Cordes \& Shannon 2008).
However, their evolutionary status is still unclear.

Known intermittent pulsars and RRATs
have characteristic ages similar to those of
``normal'' pulsars (i.e., $\ltorder10^{7}$\,yr).
However,
as we discussed in \S\ref{sec:IONS_SkyDen}, the long-duration radio transients
cannot be associated exclusively with pulsars younger
than about 1\,Gyr, otherwise their predicted surface density
will not be consistent with the
observations (\S\ref{sec:SkyDen}).

As shown in \S\ref{sec:E}, in the framework of Galactic NSs,
the duty cycle of the long-duration radio transients
is $\sim7\times 10^{-3}$. Such a small duty cycle will make it
hard to detect them as repeaters in current pulsars searches.
In addition, most pulsar searches are conducted at
low frequencies ($\ltorder 1.4$\,GHz), in which the
rate of the long-duration radio transients seems to be low.
Thus, a prediction of this model is that high frequency (say 5\,GHz)
searches should find a much larger rate of long duration radio transients.

The flat spectrum ($\alpha \gtorder 0$) of the
long duration transients is reminiscent of the
radio spectrum of magnetars in their ``active state''
(cf. Camilo et al. 2006).
Thus, a plausible model is that the long duration transients
are ancient magnetars in short lived high states.

\subsection{The distance scale to the Kida et al. transients}
\label{sec:KidaDist}

Next we
derive a physical distance to the bright events discussed by
Kida et al. (2008).
We remind the reader that these events are about a thousand
times brighter than the VLA events.
We use a modified Rayleigh test (see Fisher et al. 1987)
to compare the sky distribution of the Kida et al. (2008)
transients with that of the celestial positions of simulated NSs:
\begin{equation}
(\sum_{i=1}^{6}{\sin[b_{i}]})^{2}\equiv \Sigma^{2}b,
\label{eq:RayTest}
\end{equation}
for the sample of six radio transients found by Kida et al. (2008);
here $b$ is the Galactic latitude.

\begin{figure}
\centerline{\includegraphics[width=8.5cm]{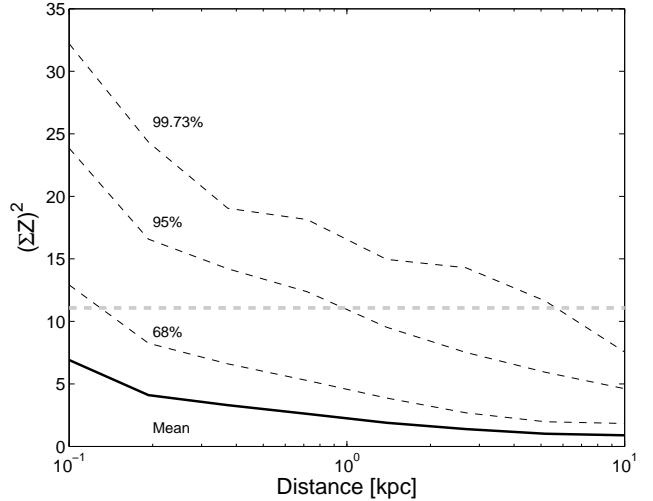}}
\caption{The sky uniformity of the Kida et al. transients
(described by $\Sigma^{2}b$; Eq.~\ref{eq:RayTest}) as a function of the
distance to the transients.
The {\it gray-thick dashed} line shows the value calculated
for the observed sample.
The {\it black} line shows the mean value calculated
for the simulated NSs, while the {\it gray-thin dashed} lines
show the upper $68$, $95$ and $99.73$ percentiles
(from bottom up) of the distribution of the
uniformity criterion $\Sigma^{2}{b}$ of the simulated NSs.
The lines fluctuates due to the limited statistics.
\label{fig:Kida_Uniformity}}
\end{figure}

Next, we selected from the Heliocentric catalog of
simulated isolated-old NSs of Ofek (2009),
six random NSs within the footprints of the Kida et al. search zone,
found up to a distance $d$ from the Sun
and calculated their $\Sigma^{2}b$.
We assume that the survey described in Kida et al. (2008)
covers the entire declination zone $\delta=32^{\circ}-42^{\circ}$
equally.
For each distance $d$,
in the range of 100\,pc to 10\,kpc,
we repeated this process 10,000 times.
In Figure~\ref{fig:Kida_Uniformity},
we show the mean expected
value, and the $68$, $95$ and $99.73$ percentiles,
for the distribution of $\Sigma^{2}b$
of simulated NSs as a function of distance.
As can be gathered from Figure~\ref{fig:Kida_Uniformity}
the typical distance to these events is below 150\,pc (900\,pc),
at the $68\%$ ($95\%$) CL.

Kida et al. reported six events in their survey
which covers about $7\%$ of the celestial sphere.
Therefore, the all-sky surface density of the Kida et al. transients
is $\gtorder30$ sources at the $95\%$ confidence.
Ofek (2009) found that the density of NSs in the solar neighborhood is
$2\times10^{-4}$\,pc$^{-3}$ assuming $10^{9}$ NS in the Galaxy
and the Arzoumanian et al. (2002) initial velocity distribution.
At small distances ($\ltorder 100$\,pc) the distribution of NSs
around the Sun is near isotropic.
Therefore, by comparing the density from the simulations
with the observed surface density of the Kida et al. (2008) events,
we put a lower limit on the distance to the Kida
et al. (2008) event, assuming they originate from Galactic NSs,
of $\gtorder30$\,pc at the 95$\%$ CL.

\section{Summary}
\label{sec:Disc}

We review several recent discoveries of radio transients
with durations between minutes to days
(Kuniyoshi et al. 2006; Bower et al. 2007; Niinuma et al. 2007; Kida et al. 2008).
We suggest that these radio transients may be generated by a single class of progenitors.
The main characteristics of these ``long duration radio transients'' are:
(i) a very high occurrence rate of
about $\sim10^{3}$\,deg$^{-2}$\,yr$^{-1}$ in the 5-GHz band;
(ii) common at intermediate Galactic latitude,
with progenitors sky surface density of $\gtorder60$\,deg$^{-2}$ at
Galactic latitude, $b\sim40^{\circ}$;
(iii) lacking any X-ray ($\gtorder10^{-13}$\,erg\,cm$^{-2}$\,s$^{-1}$),
visible light ($g<27.6$, $R<26.5$\,mag), near-IR ($K<20.4$\,mag)
and radio ($S_{5\,{\rm GHz}}\gtorder10\,\mu$Jy) counterparts;
and (iv) more abundant in the 5-GHz band as compared to that in
the 1.4-GHz band;
From the rates in the two bands we infer the spectral
index between the two bands is $\alpha \gtorder 0$
where the flux density, $f_{\nu} \propto \nu^{\alpha}$.

These events are most probably not associated
with the usual culprits like GRBs, SGRs, AGNs, SNe, flare stars, pulsars and
interacting binaries (\S\ref{sec:NotProgenitor}).
We find that several other hypothesis including 
Galactic isolated old NS; brown dwarfs
and some sort of a new kind of explosions
cannot be ruled out.
Among these, we find that the association with isolated old NSs
is especially attractive.
We explore this hypothesis in details and show that it is
consistent with the current observations.
In the framework of Galactic 
isolated old NSs
we show that:
(i) the typical distance to the Bower et al. sample (mJy events) is
between 1\,kpc and $\sim5$\,kpc;
(ii) the typical distance to the Kida et al. Jy-level events is
less than about 0.9\,kpc and more than 30\,pc at the $95\%$ CL;
and (iii) they will have a burst repetition time scale of about several months
and duty cycle of $\sim7\times 10^{-3}$.

A possible association with isolated old NS
is exciting. If correct, this may prove to be the most practical
way, so far, to find old NSs in our Galaxy and explore their
demographics.
Specifically, 
the demography of old NSs constitute an excellent probe
of the star formation history and metal enrichment of the Galaxy,
and the gravitational potential of the Galaxy.

Our analysis naturally suggests several tests that can be used
to rule out this hypothesis.
First, if indeed long-duration radio transients
are associated with isolated old NSs,
then we expect them to be distributed inhomogeneously
on the celestial sphere.
Specifically, this hypothesis predicts
that faint sub-mJy long-duration radio transients
will be at least a few times more common
in the Galactic center than in high Galactic latitude.
The exact ratio, however, depends on the distance scale
to these events.
This is illustrated in Figure~\ref{fig:NS_SkyDistribution}
which shows the expected approximate distribution
of isolated old NSs on the celestial sphere,
for the entire NS population and
NS which are at distance smaller than 1\,kpc
from the Sun.
Finally, we note that if the radio emission from such sources
is pulsating, then pulsars searches conducted at 5\,GHz
will find these objects.
The fact that such ``pulsars'' were not found in existing surveys
may be due to the fact that the majority of pulsars searches
are carried on in low frequencies in which
the rate of these transients is low.

\acknowledgments
We thank Re'em Sari, Orly Gnat, Ehud Nakar, Peter Goldreich,
Stel Phinney, Dovi Poznanski and Nat Butler
for many discussions and for Robert Becker for valuable information
regarding the strategy of the FIRST survey and information
regarding VLA\,J172059.90$+$385226.6.
Support for program number HST-GO-11104.01-A was provided by NASA through
a grant from the Space Telescope Science Institute, which is
operated by the Association of Universities for Research in
Astronomy, Incorporated, under NASA contract NAS5-26555.
A.G. acknowledges support by the Israeli Science Foundation, an
EU Seventh Framework Programme Marie Curie IRG fellowship, and the
Benoziyo Center for Astrophysics,
a research grant from the Peter and Patricia Gruber Awards,
and the William Z. and Eda Bess Novick New Scientists Fund at the
Weizmann Institute.

\end{document}